\documentclass[journal]{IEEEtran}
\usepackage{bbm}
\usepackage{tikz}
\usepackage{comment}

\usepackage{graphicx}
\ifCLASSINFOpdf
\else
\fi
\usepackage{amsmath}
\usepackage{amssymb,bm,amsthm,amsfonts}

\usepackage{booktabs}
\usepackage{makecell}

\newcommand{\V}[1]{{\bf #1}} 
\newcommand{\Vg}[1]{\bm{#1}} 

\newcounter{questctr}

\def \RR {\mathbb{R}}

\def \Ncal {\mathcal{N}}

\def \Tcal {\mathcal{T}}
\def \Ucal {\mathcal{U}}

\def \IGcal {\mathcal{I}\mathcal{G}}

\def \sumlim {\sum\limits} 

\def \fmax {f_{\text{max}}}
\def \fmin {f_{\text{min}}}

\def \Niter {N_{\text{iter}}} 

\begin{document}
%
\title{Bayesian Restoration of Audio Degraded by Low-Frequency Pulses Modeled via Gaussian Process}
%
%
%

\author{Hugo~T.~de~Carvalho,
        Flavio~R.~Avila,~\IEEEmembership{Member,~IEEE}
        and~Luiz~W.~P.~Biscainho,~\IEEEmembership{Senior Member,~IEEE}
\thanks{The authors would like to thank CAPES, CNPq, and FAPERJ agencies for funding their research work.}
\thanks{Hugo T. de Carvalho and Luiz W. P. Biscainho are with the Federal University of Rio de Janeiro (UFRJ),
Rio de Janeiro, Brazil (e-mail: hugo@dme.ufrj.br, wagner@smt.ufrj.br).}
\thanks{Flavio R. Avila is with Rio de Janeiro State University, Rio de Janeiro 20550-900,
Brazil. (e-mail: flavio.avila@uerj.br)}

}

\maketitle

\begin{abstract}
A common defect found when reproducing old vinyl and gramophone recordings with mechanical devices are the long pulses with significant low-frequency content caused by the interaction of the arm-needle system with deep scratches or even breakages on the media surface. Previous approaches to their suppression on digital counterparts of the recordings depend on a prior estimation of the pulse location, usually performed via heuristic methods. This paper proposes a novel Bayesian approach capable of jointly estimating the pulse location; interpolating the almost annihilated signal underlying the strong discontinuity that initiates the pulse; and also estimating the long pulse tail by a simple Gaussian Process, allowing its suppression from the corrupted signal. The posterior distribution for the model parameters as well for the pulse is explored via Markov-Chain Monte Carlo (MCMC) algorithms. Controlled experiments indicate that the proposed method, while requiring significantly less user intervention, achieves perceptual results similar to those of previous approaches and performs well when dealing with naturally degraded signals. 
\end{abstract}


%
\IEEEpeerreviewmaketitle

\section{Introduction}
%
%
%
%

\IEEEPARstart{A}{fairly} common degradation found in old vinyl and gramophone recordings are long pulses with significant low-frequency content produced by the nonlinear response of the arm-needle system of the playback device when it passes through a deep scratch or even a breakage on the surface of the respective media. More precisely, this degradation can be split into two complementary parts. An \emph{initial discontinuity} that arises when the needle passes exactly over the physical damage is immediately followed by the \emph{tail}, a low-frequency oscillation whose amplitude and frequency decay slowly. The discontinuity (lasting typically less than 10 ms) behaves as a high-variance noise added to the original signal that almost hides the underlying information, whereas the smooth tail (which in the most severe cases can be about 1-second long) is clearly superimposed to the original signal.

Since the physical restoration of the media is almost impossible, one must resort to numerical algorithms that process a digitized version of the degraded recording. A pioneering method tailored to tackle this type of degradation was proposed in~\cite{vaseghi-phd, vaseghi-frayling-long-pulse}: it is based on the hypothesis of similarity among the pulses present in a signal, i.e., every physical damage found during reproduction is supposed to evoke a similar response of the system, differing only in location and amplitude. These two quantities are estimated by correlating the degraded signal with prototypical pulses of a reference database. The method achieves good results when this hypothesis is valid, but its scope is limited to pulses similar to the templates present in the database. Moreover, if two or more pulses overlap, it fails.

A statistical approach capable of dealing with more general cases can be found in~\cite{godsill-rayner, godsill-tan-long-pulse}. This method assumes that both the underlying signal and the pulse can be modeled by superimposed auto-regressive (AR) processes. The original signal is then estimated by separating the two processes. Some limitations are  requiring the location of the pulse to be known and the unrealistic assumption about the AR model for the pulse.

In~\cite{esquef-et-al-long-pulse} a much simpler method is proposed, based on a nonlinear filtering technique called Two-Pass Split Window (TPSW). This filtering is employed to obtain a rough estimate of the pulse shape, which is then smoothed by a piece-wise polynomial fitting. Although this method requires less computational power, the location of the pulse must be known in advance. In~\cite{esquef-welter} the authors introduce a restoration method based on the Empirical Mode Decomposition, which decomposes a signal waveform into a set of simpler Intrinsic Mode Functions, but also requires the location of the pulse.

In general, the aforementioned methods only deal with the pulse tail, demanding also some de-clicking technique to interpolate the signal underlying the initial discontinuity. The exception is the AR separation based method~\cite{godsill-rayner, godsill-tan-long-pulse}, which assumes that the initial discontinuity is modeled by the same AR process as the tail but  with a higher excitation variance---a somewhat unrealistic hypothesis.

In~\cite{carvalho-avila-biscainho-aesbrasil-2014}, an innovative approach is introduced to jointly estimate the location of the pulse and restore the audio signal, including the excerpt underlying the initial discontinuity, that adopts a model for the pulse shape whose parameters are then estimated via Bayesian inference, by sampling from their posterior distribution. However, this method has three drawbacks, the first two being reconsidered in the present work: 1) the algorithm requires that several hyperparameters be manually tuned in order to correctly estimate the pulse tail parameters; 2) despite being capable of estimating precisely the location of the pulse, the method requires a good initialization; 3) it is not capable of dealing with superimposed pulses, because in this scenario the posterior distributions become very complicated to handle, even in the context of Markov-Chain Monte Carlo methods.

This work introduces an improvement of the method proposed in~\cite{carvalho-avila-biscainho-aesbrasil-2014} that circumvents the first two issues above: 1) the pulse tail is modeled via a Gaussian Process, requiring much less hyperparameters to be tuned; 2) an efficient initialization procedure based on~\cite{esquef-biscainho-diniz} is adopted, which provides good initial estimates of both location and duration of the initial discontinuity. But even if the posterior distribution does not increase in complexity in the case of overlapping pulses thanks to the Gaussian Process modeling of the pulse tail, the problem of estimating the underlying signal when a new initial discontinuity is superimposed to an unfinished tail still requires further investigation.


The paper is organized as follows: after this introduction, Section~\ref{sec:model-degradation} recalls the shape-based model for the pulse and introduces the proposed Gaussian Process model for the pulse tail, followed by a brief presentation of the AR-based model assumed for the underlying signal in Section~\ref{sec:model-signal}; in Section~\ref{sec:description-model} we present the complete hierarchical model on which the inference will be based, and specify prior distributions for the parameters under consideration; Section~\ref{sec:description-algorithm} briefly describes the inference 
algorithm employed, followed by the computation of the marginal likelihood in Section~\ref{sec:computation-likelihood}, and by the computation of the conditional distributions and description of the sampling procedures in Section~\ref{sec:cond-dist-sampling}; results are presented in Section~\ref{sec:results}, and conclusions are drawn in Section~\ref{sec:conclusion}.

\section{A model for the degradation}
\label{sec:model-degradation}
As already mentioned, a single pulse of the type considered in this work can be described by two contiguous parts: the initial discontinuity (when the needle passes through the physical degradation) and the tail (dumped oscillations of decaying frequency due to the non-linear response of the playback device to the impulsive excitation). These two parts of the degradation are denoted, respectively, by vectors $\V{v}_{\mathrm{d}}$ and $\V{v}_{\mathrm{t}}$. When necessary, a superscript ``s" or ``G" will be added to vector $\V{v}_{\mathrm{t}}$ in order to make explicit whether the shape-based or the Gaussian Process model is being used, respectively. There is no superscript when this distinction is not necessary.

Assuming the audio signal is processed in time frames of length $N$, denote the corresponding original and corrupted signal blocks by $\V{x}$ and $\V{y}$, respectively. In order to describe the relationship between these vectors and the degradation in vectors $\V{v}_{\mathrm{d}}$ and $\V{v}_{\mathrm{t}}$, three sets of indexes are defined: $\V{i}_0$, $\V{i}_1$ and $\V{i}_2$, indicating the time samples in $\V{y}$ that belong to the region preceding the degradation, to the initial discontinuity, and to the tail, respectively. Sub-vectors $\V{x}_0, \V{y}_0, \V{x}_1, \V{y}_1, \V{x}_2$, and $\V{y}_2$ contain the time samples corresponding to their respective sets of indexes, such that
\begin{align}
    \label{eq:dep-y-x}
    \begin{split}
        \V{y}_0 &= \V{x}_0, \\
        \V{y}_1 &= \V{x}_1 + \V{v}_{\mathrm{d}}, \\
        \V{y}_2 &= \V{x}_2 + \V{v}_{\mathrm{t}}.
    \end{split}
\end{align}
Historically, despite their simplicity, additive degradation models have been used successfully in audio restoration problems \cite{vaseghi-phd, godsill-rayner}.
By defining the set of auxiliary matrices $\V{K}$, $\V{U}_1$, and $\V{U}_2$, containing the columns of an $N \times N$ identity matrix indexed by $\V{i}_0$, $\V{i}_1$, and $\V{i}_2$, respectively, one can write
\begin{equation}
    \label{eq:x-decomposition}
	\V{x} = \V{K}\V{x}_0 + \V{U}_1\V{x}_1 + \V{U}_2\V{x}_2.
\end{equation}
In the following, the models for the initial discontinuity and the tail of the pulse are described.

\subsection{Initial discontinuity}
The initial discontinuity, stored in vector $\V{v}_{\mathrm{d}}$, can be modeled by Gaussian white noise superimposed to the underlying original signal in vector $\V{x}_1$. This part of the pulse begins at sample $n_0$ and lasts for $M$ samples, with fixed variance $\sigma_{\mathrm{d}}^2$:
\begin{align}
    \V{v}_{\mathrm{d}}(n) = r(n)[u(n - n_0) - u(n - n_0 - M)],
\end{align}
where $u(n)$ is the unit step function, $r(n) \sim \Ncal(0, \sigma_{\mathrm{d}}^2)$, and $n_0$, $M$, and $\sigma_{\mathrm{d}}^2$ are unknown. These three parameters are more concisely denoted by vector $\Vg{\theta}_{\mathrm{d}} = [n_0 ~~ M ~~ \sigma_{\mathrm{d}}^2]^T$.

\subsection{Shape-based model for the tail}
Here the model adopted in~\cite{carvalho-avila-biscainho-aesbrasil-2014} for the tail is briefly recalled. Based on~\cite{esquef-et-al-long-pulse}, it is mathematically described by 
\begin{equation}
    \label{eq:model-parametric-pulse}
    \begin{split}
        \V{v}_{\mathrm{t}}^{\mathrm{s}}(n) = V_{\mathrm{t}}{\mathrm{e}}^{-n/(f'\tau_m)}\sin\left( 2 \pi n \frac{f_n}{f'} + \phi \right) \times \\ [u(n - n_0 - M - 1)],
    \end{split}
\end{equation}
where
\begin{align}
	f_n = (\fmax - \fmin){\mathrm{e}}^{-n/(f'\tau_f)} + \fmin.
\end{align}
This model is motivated by visual inspection of pulses present in silent parts of degraded audio signals~\cite{godsill-rayner}, which exhibit a similar behavior: a decay in amplitude described by the exponential function, a decay in frequency modeled by variable $f_n$ inside the sinusoidal function. Variables $n_0$ and $M$ are defined as before, and the remaining ones are precisely defined below:
\begin{itemize}
	\item $V_{\mathrm{t}}$ is related to the tail amplitude;
	\item $f'$ is the signal sampling rate (usually 44.1 kHz);
	\item $\tau_m$ is the time constant (in seconds) associated with the pulse envelope decay;
	\item $\tau_f$ is the time constant (in seconds) associated with the pulse frequency decay;
	\item $\fmax$ and $\fmin$ are, respectively, the maximum and minimum tail oscillation frequencies (in Hz);
	\item $\phi$ is the initial phase of the pulse.
\end{itemize}

\begin{figure}[ht]
    \centering
    \includegraphics[width=8.5cm]{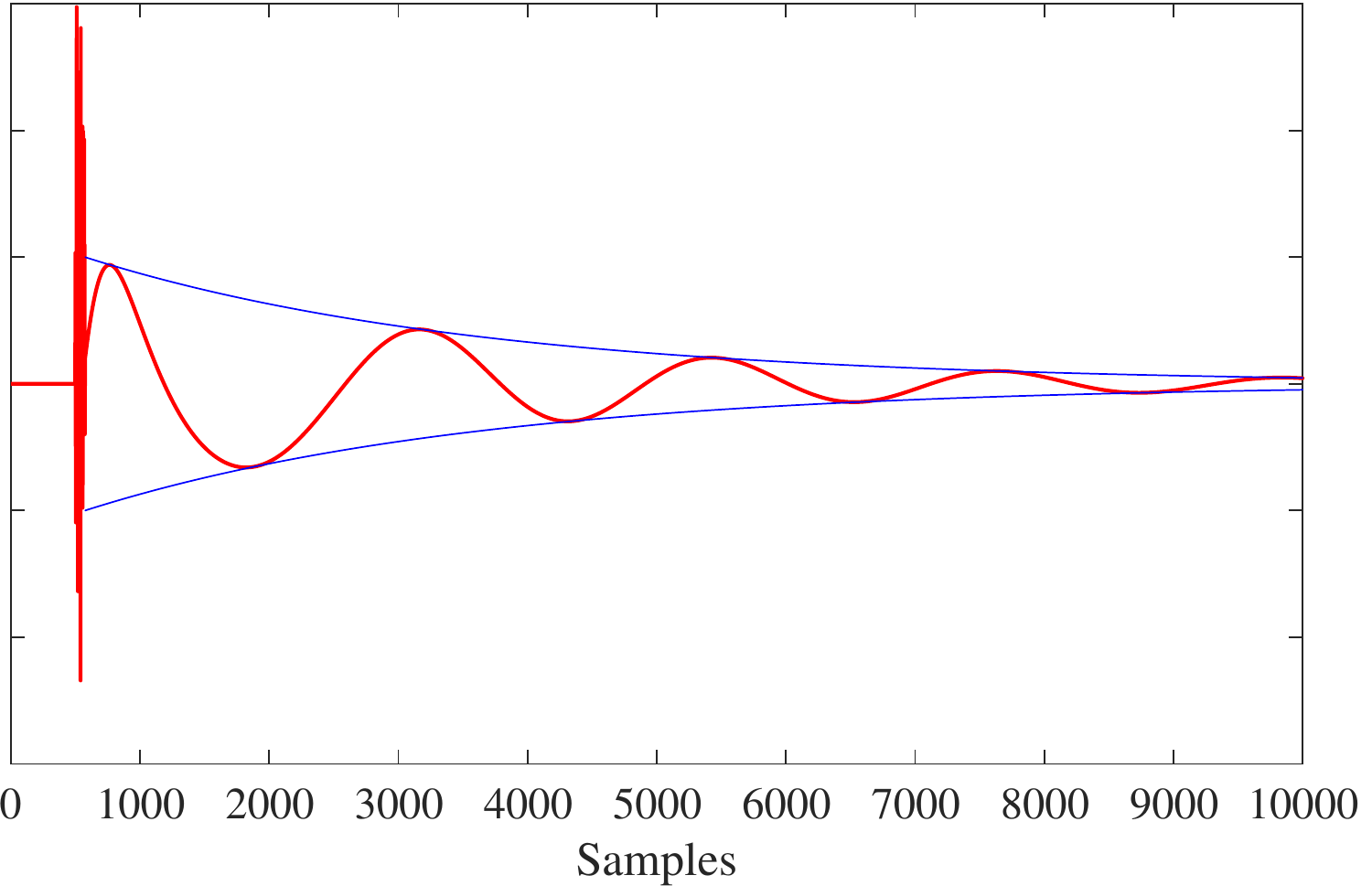}
    \caption{Model for the pulse shape following Equation~\ref{eq:model-parametric-pulse}.}
    \label{fig:parametric-pulse}
\end{figure}

All these quantities (except for $f'$) are also unknown beforehand. Let vector $\Vg{\theta}_{\mathrm{t}}^{\mathrm{s}} = [V_{\mathrm{t}} ~~ \tau_m ~~ \tau_f ~~ \fmax ~~ \fmin ~~ \phi]^T$ store such tail parameters. They can be visualized in Figure~\ref{fig:parametric-pulse}, where a prototypical pulse that follows the model above is depicted.

\subsection{Gaussian Process model for the tail}
Gaussian Processes are a widely employed technique in Statistics, Machine Learning and Linear Regression~\cite{bishop, rasmussen-williams, mackay}. Before presenting its application to the current problem, some important facts should be recalled. A stochastic process $\V{G} = \{G_t\}_{t \in \Tcal}$, indexed by some set $\Tcal$, is said to be a Gaussian Process if for any finite subset $\{t_1, \dots, t_n\} \in \Tcal$, the joint distribution of the random vector $(G_{t_1}, \dots, G_{t_n})$ is Gaussian. Note that Gaussian distributions are completely determined by their first and second order statistics, so by assuming (without loss of generality) zero mean, several features of the process are encoded in its \emph{covariance kernel}, denoted by $K(t,t')$, for $t, t' \in \Tcal$. This function describes the dependence between any two random variables (i.e. time instants) of the process, and there are several commonly used covariance kernels tailored to impose some desired structure, like stationarity, smoothness, and periodicity, among others~\cite{wilson-phd}. For a detailed discussion about Gaussian processes, see~\cite{rasmussen-williams}. 

Returning to the modeling of the pulse tail, stored in vector $\V{v}_{\mathrm{t}}$, it can be seen as a generic function superimposed to the underlying signal in vector $\V{x}_2$. This function is assumed to be much smoother than the underlying signal, and this assumption was implicitly taken into account in the design of the shape-based model. In order to give more flexibility to the pulse tail description, the deterministic function is replaced by a sample from a Gaussian Process with a squared-exponential covariance kernel given by
\begin{align}
    \label{eq:squared-exponential-covariance-kernel}
    K_{\mathrm{SE}}(t,t') = \sigma_f^2 \exp\left(-\frac{|t -t'|^2}{2\sigma_{\ell}^2}\right).
\end{align}
Parameters $\sigma_f^2$ and $\sigma_{\ell}^2$ control the amplitude of the pulse and the effective extent of the covariance kernel, respectively.

The choice of this covariance kernel to model the tail of a long pulse is arguable, since it models a stationary process, which is clearly not the case of the typical pulse tail. A more precise model would require the definition of a covariance kernel specific to this application, with some additional parameters to encode the desired behavior. The drawback of such a choice would be to excessively increase the model complexity. The good results obtained with the  squared-exponential covariance kernel consolidate our option for simplicity.

Now, in this framework, in addition to the parameters $\sigma_f^2$ and $\sigma_{\ell}^2$ of the covariance kernel, the whole vector $\V{v}_{\mathrm{t}}^{\mathrm{G}}$ containing the tail of the pulse must be estimated\footnote{Note that $\V{v}_{\mathrm{d}}$ is not included in $\Vg{\theta}_{\mathrm{d}}$, since the corresponding underlying signal $\V{x}_1$ is directly estimated instead of $\V{v}_{\mathrm{d}}$; and $\V{v}_{\mathrm{t}}^{\mathrm{s}}$ is implicitly defined by the estimated $\Vg{\theta}_{\mathrm{t}}^{\mathrm{s}}$. However, $\V{v}_{\mathrm{t}}^{\mathrm{G}}$ must be directly estimated in the Gaussian Process model, due to its non-parametric nature.}. Such quantities are assembled in vector 
\begin{align}
    \Vg{\theta}_{\mathrm{t}}^{\mathrm{G}} = [(\V{v}_{\mathrm{t}}^{\mathrm{G}})^T ~~ \sigma_f^2 ~~ \sigma_{\ell}^2]^T.
\end{align}

The estimation of parameters $\sigma_f^2$ and $\sigma_{\ell}^2$ can be performed by exploring their respective posterior distributions, in the proposed Bayesian framework. However, preliminary tests indicated that, with a proper initialization, keeping these parameters constant during the rest of the procedure is enough to reach a good pulse estimate, with the advantage of decreasing the computational cost of the algorithm. The intuition behind this observation is that with a good initialization of variables $n_0$ and $M$, guaranteed by the procedure described in Section \ref{sec:description-algorithm}, the overall shape of the pulse can be easily estimated, requiring only minor adjustments during the sampling procedure. Since parameters $\sigma_f^2$ and $\sigma_{\ell}^2$ are responsible for the overall shape of the pulse, they are not expected to change significantly during the entire procedure.


\section{A model for the underlying signal} \label{sec:model-signal}
In the context of audio restoration it is common to model the underlying original signal, here denoted by $\V{x}$, as an auto-regressive (AR) process of fixed order $P$~\cite{godsill-rayner}. More precisely, the samples of signal $\V{x}$ satisfy the following difference equation:
\begin{align}
    \label{eq:ar-model}
    x(n) = \sum_{i = 1}^{P} a_i x(n-i) + e(n),
\end{align}
where $e(n)$ is the innovation error, modeled as a white Gaussian noise with variance $\sigma_e^2$. This model is an easy way to encode the predictability of audio signals in a short scale together with some unpredictable relation between successive samples. From the signal processing viewpoint, Equation~(\ref{eq:ar-model}) describes signal $\V{x}$ as the output of an all-pole linear filter whose input is signal $\V{e} = [e(1) ~~ \dots ~~ e(N)]^T$, with transfer function given by
\begin{align}
    A(z) = \frac{1}{1 - \sumlim_{i = 1}^{P} a_i z^{-i}}.
\end{align}
Therefore, the coefficients in vector $\V{a}$ are related to the most prominent frequencies in signal $\V{x}$, through the poles of function $A(z)$.

It is important to discuss how long the block under analysis can be in order to be accurately described by an AR model. A reasonable choice is the time interval during which $\V{x}$ can be considered stationary, which for audio signals is usually accepted as around 20~ms (approximately 1,000 samples at a frequency sampling rate of 44,100~Hz, for instance). However, as mentioned before, the overall pulse duration can be much longer. The most natural way to deal with this issue would be to split signal $\V{x}$ into contiguous blocks of approximately 20~ms each and describe each of them with the corresponding AR model. However, this choice would greatly increase the complexity of the model. 

The solution proposed here is to describe 
the original signal immediately preceding the degradation and during the initial discontinuity (i.e. vector $[\V{x}_0^T ~~ \V{x}_1^T]^T$) by a single AR model of order about 40; and during the pulse tail (i.e. vector $\V{x}_2$) as white Gaussian noise---a degenerated AR model of order zero. 

This apparent oversimplification can be justified: since by hypothesis the pulse tail is much smoother than the underlying signal and consists essentially of low frequency content, in the large time-scale of the tail the signal is almost indistinguishable from white noise. Notice that this model switching makes $\V{x}$ also dependent on $n_0$ and $M$.

In contrast, a high-order AR model must be kept for the first part, since this information will be crucial to restore the virtually missed $\V{x}_1$. Here, an additional simplification can be envisaged: since estimating the coefficients of the AR model from their respective posterior distribution together with the other parameters would make the computational time very large, they are estimated beforehand from $\V{y}_0$, the region preceding the degradation, and kept constant through the whole procedure. This approximation is justifiable, since these parameters will only be used to estimate $\V{x}_1$, the signal underlying the initial discontinuity. Being $\V{y}_0 = \V{x}_0$ and contiguous to $\V{x}_1$, it is reasonable to assume they follow the same AR model.

We denote the coefficients of the AR model by $\V{a} = [a_1 ~~ \dots ~~ a_P]^T$, which, together with $\sigma_e^2$, form vector
\begin{align}
    \Vg{\theta}_x = [\V{a}^T ~~ \sigma_e^2]^T
\end{align}
containing the modeling 
parameters of the 
underlying signal.

As discussed in~\cite{godsill-rayner, carvalho-avila-biscainho-aesbrasil-2014}, by constructing matrix 
\begin{align}
    \label{eq:matrix-ar}
    \hat{\V{A}} = 
    \begin{bmatrix}
        -a_P   & \hdots & -a_1   & 1      & 0      & \hdots & 0      \\
            0  & -a_P   & \hdots & -a_1   & 1      & \ddots & \vdots \\
        \vdots & \ddots & \ddots & \ddots & \ddots & \ddots & \vdots \\
        0      & \hdots & 0      & -a_P   & \hdots & -a_1   & 1
    \end{bmatrix}
\end{align}
of size $N \times (N + P)$ and considering the distribution of the innovation error $\V{e} = \hat{\V{A}}\V{x}$ as Gaussian, by a simple change of variables one is able to write the conditional distribution for $\V{x}$, also incorporating the model switching, as
\begin{align}
    \label{eq:x-ar-gaussian}
    p(\V{x} | n_0, M, \Vg{\theta}_x) \propto \exp \left(-\frac{1}{2\sigma_e^2} {\bf x}^T{\bf A}^T{\bf A}{\bf x}\right),
\end{align}
being matrix $\V{A}$ the vertical concatenation of $\V{A}_0$, $\V{A}_1$, and $\V{A}_2$---matrices containing the columns of $\hat{\V{A}}$ indexed by $\V{i}_0$, $\V{i}_1$, and $\V{i}_2$, respectively, with all $a_j$ terms in $\V{A}_2$ set to zero, according to the simplified model adopted for $\V{x}_2$.

\section{Description of the model}
\label{sec:description-model}
We are now ready to present the whole hierarchical generative model for the degraded signal, together with the prior distributions of the parameters under consideration.

First, we can summarize the discussion up to this point about the models for the degradation and the underlying signal in the following distributions:
\begin{align}
    \label{eq:generative-model-lik}
    \begin{split}
        \V{y}_0 | n_0, \V{x}_0 &\sim \delta(\V{y}_0 - \V{x}_0) \\
        \V{y}_1 | \Vg{\theta}_\mathrm{d}, \V{x}_1 &\sim \Ncal(\V{y}_1 | \V{x}_1, \sigma_d^2 \V{I}_M) \\
        \V{y}_2 | n_0, M, \Vg{\theta}_\mathrm{t}, \V{x}_2 &\sim \delta( \V{y}_2 - (\V{x}_2 + \V{v}_{\mathrm{t}})) \\
        \V{x} | n_0, M, \Vg{\theta}_x &\sim \Ncal(\V{x} | \V{0}, \sigma_e^2(\V{A}^T\V{A})^{-1}).
    \end{split}
\end{align}
The two multi-dimensional Dirac's delta distributions for $\V{y}_0$ and $\V{y}_2$ are due to the fact that they depend deterministically on $\V{x}_0$ and $\V{x}_2 + \V{v}_{\mathrm{t}}$, respectively, as stated in Equation~(\ref{eq:dep-y-x}), whereas the distribution of $\V{y}_1$ is Gaussian as the noise superimposed on $\V{x}_1$ during the initial discontinuity. We also assume that $\V{y}_0$, $\V{y}_1$, and $\V{y}_2$ are conditionally independent given $\Vg{\theta}_{\mathrm{d}}$,  $\Vg{\theta}_{\mathrm{t}}$, and $\V{x}$. Finally, the Gaussian distribution on $\V{x}$ follows from Equation~(\ref{eq:x-ar-gaussian}).

To complete the specification of the hierarchical model, we impose prior distributions to $\Vg{\theta}_{\mathrm{d}}$ and $\Vg{\theta}_{\mathrm{t}}$, which are assumed to be prior-independent.
The components of vector $\Vg{\theta}_{\mathrm{d}}$ are assumed to be prior-independent as well. Uniform discrete distributions over the whole set of samples, denoted by $\Ucal\{\cdot | 1, \dots, N\}$, were imposed on $n_0$ and $M$. For $\sigma_{\mathrm{d}}^2$, an Inverse Gamma prior parameterized by shape $\alpha_{\mathrm{d}}$ and scale $\beta_{\mathrm{d}}$, denoted by $\IGcal(\sigma_{\mathrm{d}}^2 | \alpha_{\mathrm{d}}, \beta_{\mathrm{d}})$, was adopted; to make this prior vague, $\alpha_{\mathrm{d}} = \beta_{\mathrm{d}} = 10^{-4}$ were chosen. The prior distribution for $\Vg{\theta}_{\mathrm{t}}$, generically written as $p(\Vg{\theta}_\mathrm{t})$ for now, depends on the specific model under consideration, and will be detailed later. 
The overall prior structure can be summarized as:
\begin{align}
    \label{eq:generative-model-priors}
    \begin{split}
        n_0 &\sim \Ucal\{n_0 | 1, \dots, N\} \\
        M &\sim \Ucal\{M | 1, \dots, N\} \\
        \sigma_{\mathrm{d}}^2 | \alpha_{\mathrm{d}}, \beta_{\mathrm{d}} &\sim \IGcal(\sigma_{\mathrm{d}}^2 | \alpha_{\mathrm{d}}, \beta_{\mathrm{d}}) \\
        \Vg{\theta}_\mathrm{t} &\sim p(\Vg{\theta}_\mathrm{t}).
    \end{split}
\end{align}

\begin{figure}
    \centering
    \tikzset{every picture/.style={line width=0.75pt}} 
    \begin{tikzpicture}[x=0.75pt,y=0.75pt,yscale=-1,xscale=1]
        
        \draw   (389.93,104.02) -- (424.7,104.02) -- (424.7,206.4) -- (389.93,206.4) -- cycle ;
        \draw   (305.21,182.97) -- (337.1,182.97) -- (337.1,241.4) -- (305.21,241.4) -- cycle ;
        \draw   (150.97,151.38) -- (226.97,151.38) -- (226.97,232.4) -- (150.97,232.4) -- cycle ;
        \draw   (243,151.2) -- (288.7,151.2) -- (288.7,337.4) -- (243,337.4) -- cycle ;
        \draw    (186.7,122.2) -- (186.7,148.2) ;
        \draw    (264.7,122.2) -- (264.7,147.2) ;
        \draw    (320.7,153.2) -- (320.7,178.2) ;
        \draw    (405.7,73.2) -- (405.7,98.2) ;
        \draw  [dash pattern={on 0.84pt off 2.51pt}]  (220.7,69.2) -- (196.7,94.2) ;
        \draw  [dash pattern={on 0.84pt off 2.51pt}]  (241.7,70.2) -- (265.7,95.2) ;
        
        \draw (322,18.36) node    {$\V{y}$};
        
        \draw (322,75.31) node    {$\V{x}$};
        
        \draw (407,58.74) node    {$\Vg{\theta}_{\mathrm{d}}$};
        \draw (407,118.38) node    {$n_0$};
        \draw (407,153.58) node    {$M$};
        \draw (407,186.72) node    {$\sigma_{\mathrm{d}}^2$};
        \draw (381,237.35) node    {$\alpha_{\mathrm{d}}$};
        \draw (433,237.35) node    {$\beta_{\mathrm{d}}$};
        
        \draw (322,140.55) node    {$\Vg{\theta}_x$};
        \draw (322,224.67) node    {$\sigma_e^2$};
        \draw (322,194.18) node    {$\V{a}$};
        
        \draw (232,56.67) node    {$\Vg{\theta}_{\mathrm{t}}$};
        
        \draw (190,106.38) node    {$\Vg{\theta}_{\mathrm{t}}^{\mathrm{G}}$};
        \draw (190,165.08) node    {$\V{v}_{\mathrm{t}}^{\mathrm{G}}$};
        \draw (164,217.82) node    {$\sigma_f^2$};
        \draw (216,216.86) node    {$\sigma_{\ell}^2$};
        
        \draw (265,107.45) node    {$\Vg{\theta}_{\mathrm{t}}^{\mathrm{s}}$};
        \draw (265,163.69) node    {$V_{\mathrm{t}}$};
        \draw (265,191.83) node    {$\tau_m$};
        \draw (265,226.07) node    {$\tau_f$};
        \draw (265,261.24) node    {$\fmax$};
        \draw (265,294.27) node    {$\fmin$};
        \draw (265,326.44) node    {$\phi$};
        
        \draw    (322.01,63.31) -- (322.04,32.36) ;
        \draw [shift={(322.04,30.36)}, rotate = 450.05] [color={rgb, 255:red, 0; green, 0; blue, 0 }  ][line width=0.75]    (10.93,-3.29) .. controls (6.95,-1.4) and (3.31,-0.3) .. (0,0) .. controls (3.31,0.3) and (6.95,1.4) .. (10.93,3.29)   ;
        \draw    (395.56,53.07) -- (333.36,23.7) ;
        \draw [shift={(331.55,22.84)}, rotate = 385.28] [color={rgb, 255:red, 0; green, 0; blue, 0 }  ][line width=0.75]    (10.93,-3.29) .. controls (6.95,-1.4) and (3.31,-0.3) .. (0,0) .. controls (3.31,0.3) and (6.95,1.4) .. (10.93,3.29)   ;
        \draw    (243.23,51.98) -- (310.71,23.19) ;
        \draw [shift={(312.55,22.41)}, rotate = 516.9] [color={rgb, 255:red, 0; green, 0; blue, 0 }  ][line width=0.75]    (10.93,-3.29) .. controls (6.95,-1.4) and (3.31,-0.3) .. (0,0) .. controls (3.31,0.3) and (6.95,1.4) .. (10.93,3.29)   ;
        \draw    (395.01,110.37) -- (333.3,80.74) ;
        \draw [shift={(331.5,79.87)}, rotate = 385.65] [color={rgb, 255:red, 0; green, 0; blue, 0 }  ][line width=0.75]    (10.93,-3.29) .. controls (6.95,-1.4) and (3.31,-0.3) .. (0,0) .. controls (3.31,0.3) and (6.95,1.4) .. (10.93,3.29)   ;
        \draw    (395.61,139.17) -- (333.01,84.86) ;
        \draw [shift={(331.5,83.55)}, rotate = 400.94] [color={rgb, 255:red, 0; green, 0; blue, 0 }  ][line width=0.75]    (10.93,-3.29) .. controls (6.95,-1.4) and (3.31,-0.3) .. (0,0) .. controls (3.31,0.3) and (6.95,1.4) .. (10.93,3.29)   ;
        \draw    (322,127.55) -- (322,89.31) ;
        \draw [shift={(322,87.31)}, rotate = 450] [color={rgb, 255:red, 0; green, 0; blue, 0 }  ][line width=0.75]    (10.93,-3.29) .. controls (6.95,-1.4) and (3.31,-0.3) .. (0,0) .. controls (3.31,0.3) and (6.95,1.4) .. (10.93,3.29)   ;
        \draw    (387.41,224.35) -- (400.03,197.04) ;
        \draw [shift={(400.86,195.22)}, rotate = 474.79] [color={rgb, 255:red, 0; green, 0; blue, 0 }  ][line width=0.75]    (10.93,-3.29) .. controls (6.95,-1.4) and (3.31,-0.3) .. (0,0) .. controls (3.31,0.3) and (6.95,1.4) .. (10.93,3.29)   ;
        \draw    (428.53,223.32) -- (415.58,197.01) ;
        \draw [shift={(414.7,195.22)}, rotate = 423.8] [color={rgb, 255:red, 0; green, 0; blue, 0 }  ][line width=0.75]    (10.93,-3.29) .. controls (6.95,-1.4) and (3.31,-0.3) .. (0,0) .. controls (3.31,0.3) and (6.95,1.4) .. (10.93,3.29)   ;
        \draw    (171.24,202.32) -- (182.39,179.87) ;
        \draw [shift={(183.28,178.08)}, rotate = 476.42] [color={rgb, 255:red, 0; green, 0; blue, 0 }  ][line width=0.75]    (10.93,-3.29) .. controls (6.95,-1.4) and (3.31,-0.3) .. (0,0) .. controls (3.31,0.3) and (6.95,1.4) .. (10.93,3.29)   ;
        \draw    (207.87,202.36) -- (197.52,179.9) ;
        \draw [shift={(196.69,178.08)}, rotate = 425.27] [color={rgb, 255:red, 0; green, 0; blue, 0 }  ][line width=0.75]    (10.93,-3.29) .. controls (6.95,-1.4) and (3.31,-0.3) .. (0,0) .. controls (3.31,0.3) and (6.95,1.4) .. (10.93,3.29)   ;
    \end{tikzpicture}
    \caption{Graphical dependence structure in the proposed model. Arrows, solid lines and dotted lines indicate, respectively: direct dependence, display of vector components, and both modeling possibilities for $\Vg{\theta}_\mathrm{t}$.}
    \label{fig:hierarchical-model}
\end{figure}
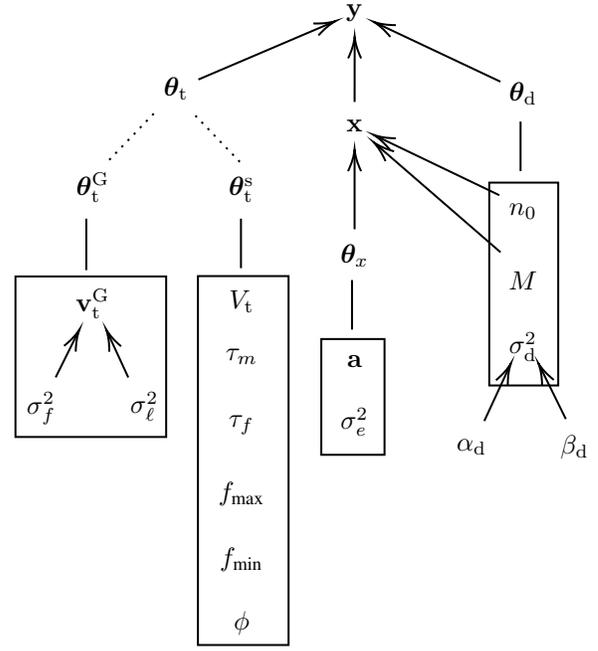

The dependence relations between signals and parameters are shown as a graphical model in Figure \ref{fig:hierarchical-model}, where arrows indicate direct dependencies; solid lines denote the splitting of a vector into its components; and the dotted lines after vector $\Vg{\theta}_\mathrm{t}$ refer to the two possible models for the pulse tail.

\subsection{Prior distribution for $\Vg{\theta}_\mathrm{t}^{\mathrm{s}}$}
The dependence between $\V{y}_2$ and $\Vg{\theta}_\mathrm{t}^{\mathrm{s}}$ is quite complex when considering the shape-based model for the pulse tail. For this reason, we chose to impose a simple non-informative prior to the components of $\Vg{\theta}_\mathrm{t}^{\mathrm{s}}$. After assuming their mutual prior-independence, improper uniform priors in their respective domains were adopted:
\begin{equation}
    p(\Vg{\theta}_\mathrm{t}^{\mathrm{s}}) \propto \mathbbm{1}(\tau_m > 0) \mathbbm{1}(\tau_f > 0) \mathbbm{1}(\fmax > 0) \mathbbm{1}(\fmin > 0),
\end{equation}
where $\mathbbm{1}$ denotes an indicator function. Note that variables $V_\mathrm{t}$ and $\phi$ can assume any real value. Even if this choice of prior distribution prioritize simplicity over accuracy, it was verified in preliminary tests that improper priors does not negatively impact the estimation procedure.

\subsection{Prior distribution for $\Vg{\theta}_\mathrm{t}^{\mathrm{G}}$}
Since in this scenario we are assuming that the pulse tail, stored in vector $\V{v}_{\mathrm{t}}^{\mathrm{G}}$, is well described by a Gaussian Process with a squared-exponential covariance kernel, its prior distribution is given by
\begin{equation}
    \label{eq:gp-prior-tail}
    \V{v}_{\mathrm{t}}^{\mathrm{G}} | \sigma_f^2, \sigma_{\ell}^2 \sim \Ncal(\V{v}_{\mathrm{t}}^{\mathrm{G}} | \V{0}, \V{C}),
\end{equation}
where matrix $\V{C}$ is the Gram matrix of the covariance kernel function, computed from Equation~(\ref{eq:squared-exponential-covariance-kernel}).

No prior distributions are assigned to $\sigma_f^2$ and $\sigma_{\ell}^2$, since (as stated in Section \ref{sec:model-degradation}) they will be estimated beforehand and kept constant during the rest of the estimation procedure.

\section{Description of the algorithm}
\label{sec:description-algorithm}
Having defined the model in Equations (\ref{eq:generative-model-lik}) and (\ref{eq:generative-model-priors}), we now describe the inference strategy adopted. First, the initialization procedure for $n_0$ and $M$ is presented, followed by a description of the estimation of $\Vg{\theta}_x$, $\sigma_f^2$ and $\sigma_{\ell}^2$. This section closes with an overview of the sampling algorithm.

\subsection{Initialization procedure for $n_0$ and $M$}
Previous experiments indicated that an accurate initialization of variables $n_0$ and $M$ was important for the sampling procedure. Performing this initialization by hand (i.e., choosing the respective initial values by a visual inspection of the signal) may not be practical, and some method for automatically doing this task was required. An adaptation of the method for detection of long pulses in audio signals proposed in~\cite{esquef-biscainho-diniz} was able to provide a quite accurate first estimate of the desired variables.

The initialization procedure looks for sudden high-amplitude impulses in time (typical of the initial discontinuity of the pulse), whose energy splits over the whole spectrum of a time frame around it. More precisely, the corrupted signal is split in contiguous blocks of length $L$ with an overlap of 50\% between adjacent blocks, and the discrete Fourier Transform (DFT)~\cite{oppenheim-schafer-sig-prog} of each block is computed. Denote the DFT of block $b$ by $\hat{\V{y}}_b$, for $b = 1, \dots B$. Since typically the power of the audio spectrum is concentrated below some medium-high frequency, it is more convenient to look for unusual information in vectors $\hat{\V{y}}_b$ above some frequency defined by the user. Denote this cut-off frequency by $f_\mathrm{co}$ and its respective frequency bin by $\alpha_\mathrm{co}$. Define the function
\begin{equation}
	\mu(b) = \frac{1}{\beta - \alpha_\mathrm{co} + 1}\sum_{k = \alpha_\mathrm{co}}^{\beta} |\hat{\V{y}}_b(k)|,
\end{equation}
where $\beta$ indexes the last bin in the DFT. This function is an arithmetic mean of the high-frequency content in block $b$, starting at frequency $f_\mathrm{co}$. It is expected to reach a high value when the initial discontinuity of a pulse occurs in a given block of the degraded signal. 

However, as reported in~\cite{esquef-biscainho-diniz}, if the considered signal exhibits a broad dynamic range with substantial high-frequency content (for example, brass or percussive instruments), the value of $\mu$ can be high without necessarily implying the presence of long pulses.
In order to circumvent this issue, here a median filter is applied to function $\mu$. As it is known in the literature, the median filter is capable of removing local occurrences of unusual values within a sequence, and is widely used in Image Processing as a tool to remove impulsive noise while preserving edges~\cite{jain}. 
The overall procedure is described below:
\begin{itemize}
    \item Pad function $\mu$ with $\left \lfloor{c/2}\right \rfloor$ zeros before its first and after its last samples, respectively.
	\item Define a new function $\mu_m$ resulting of median filtering $\mu$ with an odd-sized window of length $c$, that is, by replacing each value of $\mu$ by the empirical median of the $c$ values around it.
	\item Define function $\Delta \mu(b)$ as the difference between $\mu$ and $\mu_m$ normalized by its highest value, that is,
	\begin{equation}
	    \label{eq:init-proc-norm}
		\Delta \mu(b) = \frac{\mu(b) - \mu_m(b)}{\max\limits_{b'} [\mu(b') - \mu_m(b')]},
	\end{equation}
	for $b = 1, \dots, B$. This ensures that the maximum absolute value of $\Delta \mu$ is one, thus allowing an easier definition of the threshold specified below.
	\item Define a threshold $\xi$ such that block $b^*$ is considered corrupted by the initial discontinuity of a long pulse if $|\Delta \mu(b^*)| \geq \xi$.
\end{itemize}

This procedure defines a set of causally ordered candidate blocks $b_1^*, \dots, b_M^*$. Each contiguous subset of blocks $b_i^*,\dots,b_j^*$ is attributed to the initial discontinuity of a long pulse, for which $n_0$ is chosen as the first time sample of block $b_i^*$, and $M$ as the gap size between the last time sample of block $b_j^*$ and $n_0$.


In summary, in the initialization step the values of $L$, $f_\mathrm{co}$, and $\xi$ are left to the user's choice. However, in order to reach a useful initialization, the value of $L$, representing the length of each block prior to the computation of its DFT, must be carefully chosen. For simplicity, all the block lengths here refer to a signal sampled at the frequency rate of 44,1 kHz. In~\cite{esquef-biscainho-diniz} the authors adopted $L = 2048$ ($\approx$ 46~ms). Since the portion of the signal being analyzed that contains the initial discontinuity and the pulse tail typically spans around 10,000 samples ($\approx$ 226~ms), this choice of $L$ would imply a crude time resolution, thus providing loose initial estimates for $n_0$ and $M$. In order to increase time resolution, a value of $L$ between 16 and 64 is suggested here. Experiments indicate that when the degraded signal is also contaminated with broadband additive noise, at low SNR, values of $L$ near the upper-range allow for better estimates. Besides $L\in\{16,32,64\}$, $\xi$ around $0.3$ is recommended, and $f_{\mathrm{co}}=\frac{f'}{2}$ is a good starting point.

It is important to remark that this initialization procedure is specifically tailored to be applied on signals effectively containing at least one pulse, mainly due to the normalization in Equation~(\ref{eq:init-proc-norm}). Indeed, tests indicate that with an undistorted signal, any higher-frequency content is likely to be detected as a pulse with this procedure. This issue can be circumvented by dropping the normalization in Equation~(\ref{eq:init-proc-norm}); however, it makes the definition of the threshold $\xi$ more subtle, requiring a deeper examination of the signal under consideration. Since the existence of a pulse is easily confirmed by informal listening of the signal, we have opted for keeping the normalization.

\subsection{Estimation of $\Vg{\theta}_x$, $\sigma_f^2$ and $\sigma_{\ell}^2$}
After initialization of variables $n_0$ and $M$, we can consider the excerpts $\V{y}_0^{(0)}$ and $\V{y}_2^{(0)}$ of the degraded signal, before and after the initial estimate of the initial discontinuity respectively, from which $\Vg{\theta}_x$, $\sigma_f^2$ and $\sigma_{\ell}^2$ will be estimated and kept constant during the execution of the sampling algorithm.

Since $\V{y}_0^{(0)}$ is assumed to be uncorrupted, the parameters of the AR model, stored in vector $\Vg{\theta}_x$, are estimated from this portion of the signal by the covariance method \cite{cryer-chan}. As for the Gaussian Process model for the pulse tail, $\V{v}_{\mathrm{t}}^{\mathrm{G}}$, along with parameters $\sigma_f^2$ and $\sigma_{\ell}^2$, are initialized by fitting a Gaussian Process with the chosen squared-exponential kernel to $\V{y}_2^{(0)}$, via a standard maximum likelihood procedure~\cite{rasmussen-williams}.

Note that in the shape-based model for the pulse tail the initialization of $\Vg{\theta}_{\mathrm{t}}^{\mathrm{s}}$ is done by the user. In Section~\ref{sec:results} a suggested set of parameters produces a shape that resembles an actual pulse. 


\subsection{Sampling algorithm}
At this point, the goal is to recover the underlying signal $\V{x}$ from the observed degraded signal $\V{y}$. This can be achieved by estimating the auxiliary quantities stored in vectors $\Vg{\theta}_{\mathrm{d}}$ and $\Vg{\theta}_{\mathrm{t}}$---the latter related to either the shape-based or the Gaussian Process model for the pulse tail, 
as well as the almost missing signal $\V{x}_1$ underlying the initial discontinuity. After some manipulation of Equation~(\ref{eq:generative-model-lik}), it is possible to obtain the likelihood $p(\V{y} | \V{x}, \Vg{\theta}_x, \Vg{\theta}_{\mathrm{d}}, \Vg{\theta}_{\mathrm{t}})$. Bayes' Theorem leads to the posterior distribution of the desired quantities
\begin{align}
    p(\V{x}, \Vg{\theta}_{\mathrm{d}}, \Vg{\theta}_{\mathrm{t}} | \Vg{\theta}_x, \V{y}).
\end{align}
Since this expression is analytically intractable, this distribution is sampled from via the Gibbs sampler~\cite{geman-geman, robert-casella-small, robert-casella-big} (eventually with some Metropolis steps if the corresponding conditional distribution is not from a known family of distributions), implemented as follows:
\begin{enumerate}
    \item [a) ] Initialize values $n_0^{(0)}$, $M^{(0)}$, $\sigma_{\mathrm{d}}^{2^{(0)}}$, $\Vg{\theta}_{\mathrm{t}}^{(0)}$ and $\V{x}^{(0)}$.
    \item [b) ] For $k$ from 1 to $\Niter$:
    \begin{enumerate}
        \item [i) ] Sample $n_0^{(k)}$ and $M^{(k)}$ from distribution $$p(n_0, M | \sigma_{\mathrm{d}}^{2^{(k-1)}}, \Vg{\theta}_{\mathrm{t}}^{(k-1)}, \V{x}^{(k-1)}, \Vg{\theta}_x, \V{y}).$$
		\item [ii) ] Sample  $\Vg{\theta}_{\mathrm{t}}^{(k)}$ and $\V{x}^{(k)}$ from distribution $$p(\Vg{\theta}_{\mathrm{t}}, \V{x} | n_0^{(k)}, M^{(k)}, \sigma_{\mathrm{d}}^{2^{(k-1)}}, \Vg{\theta}_x, \V{y}).$$
		\item [iii) ] Sample $\sigma_{\mathrm{d}}^{2^{(k)}}$ from distribution $$p(\sigma_{\mathrm{d}}^2 | n_0^{(k)}, M^{(k)}, \Vg{\theta}_{\mathrm{t}}^{(k)}, \V{x}^{(k)}, \Vg{\theta}_x, \V{y}).$$
    \end{enumerate}
\end{enumerate}
Variables $n_0$ and $M$ are jointly sampled, since block Gibbs sampling is empirically known to improve the convergence speed of the algorithm~\cite{givens-hoeting-computational-statistics}. The joint sampling of $\Vg{\theta}_{\mathrm{t}}$ and $\V{x}$ is adopted for the same reason, in addition to the fact that their joint distribution can be rewritten, via Bayes' Theorem and recalling the prior independence between $\Vg{\theta}_{\mathrm{d}}$ and $\Vg{\theta}_{\mathrm{t}}$, in the following form,  which will simplify some computations in Section \ref{sec:computation-likelihood}:
\begin{equation}
    \label{eq:decomp-dist-thetat-x}
    \begin{split}
        \!\!p(\Vg{\theta}_{\mathrm{t}},& \V{x} | \Vg{\theta}_{\mathrm{d}}, \Vg{\theta}_x, \V{y}) = p(\Vg{\theta}_{\mathrm{t}} | \Vg{\theta}_{\mathrm{d}}, \Vg{\theta}_x, \V{y})p(\V{x} | \Vg{\theta}_{\mathrm{t}}, \Vg{\theta}_{\mathrm{d}}, \Vg{\theta}_x, \V{y}) \propto \\
        & [p(\V{y} | \Vg{\theta}_{\mathrm{t}}, \Vg{\theta}_{\mathrm{d}}, \Vg{\theta}_x)p(\Vg{\theta}_{\mathrm{t}})] p(\V{x} | \Vg{\theta}_{\mathrm{t}}, \Vg{\theta}_{\mathrm{d}}, \Vg{\theta}_x, \V{y}).
    \end{split}
\end{equation}

The order of the sampling is quite arbitrary, since it does not affect the convergence properties of the algorithm~\cite{givens-hoeting-computational-statistics}. This particular order was implemented since it seems more natural to first sample $n_0$ and $M$, the pulse location variables, and then sample the other variables. Finally, the mean of the posterior distribution, estimated by averaging the samples obtained after the burn-in time, is used to perform the restoration procedure.

Notice that the sampling of $\Vg{\theta}_{\mathrm{t}}$ depends on the considered model. For the shape-based model, the posterior distribution $p(\Vg{\theta}_{\mathrm{t}}^{\mathrm{s}} | \V{x}, \Vg{\theta}_{\mathrm{d}}, \Vg{\theta}_x, \V{y})$ is very complicated and does not belong to any known family of distributions; therefore, in this case, a Metropolis-Hastings step is performed within the Gibbs sampler---for more details, see~\cite{carvalho-avila-biscainho-aesbrasil-2014}. Hereafter, the focus will be on the computation of the Gaussian Process model only, detailing the derivation of each conditional distribution of interest. For the sake of completeness, the corresponding  results for the shape-based model are also reported.

\section{Computation of the marginal likelihood}
\label{sec:computation-likelihood}
This section develops the marginal likelihood $p(\V{y} | \Vg{\theta}_{\mathrm{t}}, \Vg{\theta}_{\mathrm{d}}, \Vg{\theta}_x)$, which will be needed later.

Firstly, two important results about Gaussian distributions must be recalled~\cite{godsill-rayner}.
\begin{itemize}
    \item Consider an integral of the form
    \begin{align}
        I = \int_{\RR^D} \exp \left\{-\frac{1}{2}(a + \V{b}^T\V{z} + \V{z}^T\V{C}\V{z})\right\}\mathrm{d}\V{z}.
    \end{align}
    Since the term inside the exponential function is a quadratic form on $\V{z}$, one could complete the squares and compare the obtained expression with the probability density function of a multivariate Gaussian. One can then prove that~\cite{godsill-rayner}
    \begin{align}
        I = \frac{(2\pi)^{D/2}}{\det(\V{C})^{1/2}}\exp\left\{-\frac{1}{2}\left(a - \frac{\V{b}^T\V{C}^{-1}\V{b}}{4}\right)\right\}.
    \end{align}
        
    \item Now, consider the product of two multivariate Gaussian probability density functions, that is,
    \begin{align}
        f(\V{z}) = f_1(\V{z})f_2(\V{z}),
    \end{align}
    where $f_1(\V{z}) = \Ncal(\V{z} | \Vg{\mu}_1, \Vg{\Sigma}_1)$ and $f_2(\V{z}) = \Ncal(\V{z} | \Vg{\mu}_2, \Vg{\Sigma}_2)$. Also after completing the squares inside the exponentials, one is able to prove that $f(\V{z})$ is also the probability density function of a multivariate Gaussian distribution, but with covariance matrix given by $\Vg{\Sigma} = (\Vg{\Sigma}_1^{-1} + \Vg{\Sigma}_2^{-1})^{-1}$ and mean $\Vg{\mu} = \Vg{\Sigma}^{-1}(\Vg{\Sigma}_1^{-1}\Vg{\mu}_1 + \Vg{\Sigma}_2^{-1}\Vg{\mu}_2)$.
\end{itemize}

Proceeding to the derivation of the marginal likelihood, it can be rewritten as
\begin{align}
    \label{eq:int-likelihood-step}
    \begin{split}
	    p(\V{y} |& \Vg{\theta}_{\mathrm{t}}, \Vg{\theta}_{\mathrm{d}}, \Vg{\theta}_x) = \int_{\RR^N} p(\V{x}, \V{y} | \Vg{\theta}_{\mathrm{t}}, \Vg{\theta}_{\mathrm{d}}, \Vg{\theta}_x)~\mathrm{d}\V{x} =\\
	    & \int_{\RR^N} p(\V{x} | \Vg{\theta}_{\mathrm{t}}, \Vg{\theta}_{\mathrm{d}}, \Vg{\theta}_x) p(\V{y} | \V{x}, \Vg{\theta}_{\mathrm{t}}, \Vg{\theta}_{\mathrm{d}}, \Vg{\theta}_x)~\mathrm{d}\V{x}.
	\end{split}
\end{align}

Note that the dependence of $\V{x}$ on $\Vg{\theta}_{\mathrm{t}}$ and $\sigma_{\mathrm{d}}^2$ can be dropped, and from Equation~(\ref{eq:x-ar-gaussian}), $p(\V{x} | n_0, M, \Vg{\theta}_x)$ is Gaussian with mean $\V{0}$ and covariance matrix $\sigma_e^2(\V{A}^T\V{A})^{-1}$. From Equation~(\ref{eq:generative-model-lik}) and recalling the conditional independence of $\V{y}_0$, $\V{y}_1$ and $\V{y}_2$ given $\Vg{\theta}_{\mathrm{d}}$, $\Vg{\theta}_{\mathrm{t}}$ and $\V{x}$, we can obtain
\begin{align}
    \label{eq:likelihood-y}
    \begin{split}
	    p(\V{y} |& \V{x}, \Vg{\theta}_{\mathrm{t}}, \Vg{\theta}_{\mathrm{d}}, \Vg{\theta}_x) =\\
	    &\delta(\V{y}_0 - \V{x}_0) \Ncal(\V{y}_1 | \V{x}_1, \sigma_{\mathrm{d}}^2\V{I}_M) \delta(\V{y}_2 - (\V{x}_2 + \V{v}_{\mathrm{t}})),
    \end{split}
\end{align}
and therefore
\begin{align}
    \label{eq:prod-int}
    \begin{split}
	    p(\V{x},&\V{y} | \Vg{\theta}_{\mathrm{t}}, \Vg{\theta}_{\mathrm{d}}, \Vg{\theta}_x) =
	    \Ncal(\V{x} | \V{0}, \sigma_e^2(\V{A}^T\V{A})^{-1}) \times \\ &[\delta(\V{y}_0 - \V{x}_0)\Ncal(\V{y}_1 | \V{x}_1, \sigma_{\mathrm{d}}^2\V{I}_M)\delta(\V{y}_2 - (\V{x}_2 + \V{v}_{\mathrm{t}}))].
	\end{split}
\end{align}

Notice that the integral in Equation~(\ref{eq:int-likelihood-step}) is calculated with respect to $\V{x}$, and the second Gaussian in Equation~(\ref{eq:prod-int}) depends on $\V{x}_1$ only via its mean. In order to make explicit the dependence on $\V{x}_1$, one can use the symmetry of the Gaussian distribution and the fact that $\Ncal(\V{y}_1 | \V{x}_1, \sigma_{\mathrm{d}}^2\V{I}_M) = \Ncal(\V{x}_1 | \V{y}_1, \sigma_{\mathrm{d}}^2\V{I}_M)$ (i.e., both PDFs have the same formula). Using the decomposition of $\V{x}$ given in Equation~(\ref{eq:x-decomposition}):
\begin{align}
    \label{eq:joint-dist-x-y}
    \begin{split}
	    p(\V{x},& \V{y} | \Vg{\theta}_{\mathrm{t}}, \Vg{\theta}_{\mathrm{d}}, \Vg{\theta}_x) = \\ &\Ncal(\V{K}\V{x}_0 + \V{U}_1\V{x}_1 + \V{U}_2\V{x}_2 | \V{0}, \sigma_e^2(\V{A}^T\V{A})^{-1}) \times \\
	    & [\delta(\V{y}_0 - \V{x}_0) N(\V{x}_1 | \V{y}_1, \sigma_{\mathrm{d}}^2\V{I}_M)  \delta(\V{y}_2 - (\V{x}_2 + \V{v}_{\mathrm{t}}))],
	\end{split}
\end{align}
and finally,
\begin{align} \label{eq:likelihood-parameters-long-pulse-v2.0}
    \begin{split}
	    &p(\V{y} | \Vg{\theta}_{\mathrm{t}}, \Vg{\theta}_{\mathrm{d}}, \Vg{\theta}_x) = \\
	    &\!\int_{\RR^N} \!\!\! \Ncal(\V{K}\V{x}_0 + \V{U}_1\V{x}_1 + \V{U}_2\V{x}_2 | \V{0}, \sigma_e^2(\V{A}^T\V{A})^{-1}) \times  \\ &~~[\delta(\V{y}_0 -\V{x}_0)  \Ncal(\V{x}_1 | \V{y}_1, \sigma_{\mathrm{d}}^2\V{I}_M) \delta(\V{y}_2 - (\V{x}_2 \! + \! \V{v}_{\mathrm{t}}))]\mathrm{d}\V{x} = \\ &\!\int_{\RR^M} \!\!\! \Ncal(\V{K}\V{y}_0 \! + \! \V{U}_1\V{x}_1 \! + \! \V{U}_2(\V{y}_2 - \V{v}_{\mathrm{t}}) | \V{0}, \sigma_e^2(\V{A}^T\V{A})^{-1}) \times \\ &~~\Ncal(\V{x}_1 | \V{y}_1, \sigma_{\mathrm{d}}^2\V{I}_M)~\mathrm{d}\V{x}_1.
	\end{split}
\end{align}
The integral can be computed by using the Gaussian PDFs along with the results in the beginning of this Section with $\V{z} = \V{x}_1$:
\begin{equation}
    \label{eq:partial-likelihood-y}
	p(\V{y} | \Vg{\theta}_{\mathrm{t}}, \Vg{\theta}_{\mathrm{d}}, \Vg{\theta}_x) = \frac{\lambda^M}{(2\pi \sigma_e^2)^{\frac{N - P}{2}}\det(\Vg{\Phi})}\exp\left\{-\frac{E_{\text{min}}}{2\sigma_e^2}\right\},
\end{equation}
where
\begin{align}     
    \label{eq:patial-likelihood-y-cont}
	\begin{split}
	    E_{\text{min}} &= \lambda^2\V{y}_1^T\V{y}_1 + \V{z}^T\begin{bmatrix}\V{A}_0^T \\ 
	    \V{A}_2^T\end{bmatrix}[\V{A}_0 ~ \V{A}_2]\V{z} + (\V{x}_1^{\text{MAP}})^T\Vg{\Theta}, \\
	    \V{z} &= \begin{bmatrix}\V{y}_0 \\ \V{y}_2 - \V{v}_{\mathrm{t}} \end{bmatrix}, \\
	    \V{x}_1^{\text{MAP}} &= \Vg{\Phi}^{-1}\Vg{\Theta}, \\
	    \Vg{\Phi} &= \lambda\V{I}_M + \V{A}_1^T\V{A}_1, \\
	    \Vg{\Theta} &= \lambda\V{y}_1 - \V{A}_1^T[\V{A}_0 ~ \V{A}_2]\V{z}, \\
	    \lambda &= \sigma_e^2/\sigma_{\mathrm{d}}^2.
	\end{split}
\end{align}

This expression can be simplified by noting that $\lambda$ is likely to be very small, since $\sigma_{\mathrm{d}}^2$ is usually several orders of magnitude greater than $\sigma_e^2$. In the argument of the exponential in Equation~(\ref{eq:partial-likelihood-y}) this quantity multiplies $\V{y}_1$, whose entries do not typically exceed $3\sigma_{\mathrm{d}}$ beyond the underlying signal, due to their Gaussian distribution. One can then ignore all terms involving $\lambda$ inside the exponential in Equation~(\ref{eq:partial-likelihood-y}), which becomes proportional to $\exp(-\frac{1}{2}\V{z}^T\V{R}\V{z})$, where
\begin{equation}
    \label{eq:partial-likelihood-y-simp-R}
	\V{R} = \frac{1}{\sigma_e^2}\begin{bmatrix}\V{A}_0^T \\ 
	\V{A}_2^T\end{bmatrix}\V{S}[\V{A}_0 ~ \V{A}_2],
\end{equation}
with
\begin{equation}
    \label{eq:partial-likelihood-y-simp-S}
	\V{S} = \V{I}_{N-P} - \V{A}_1(\V{A}^{-1}\V{A})^{-1}\V{A}_1^T.
\end{equation}

In several steps of the algorithm the distribution $p(\V{x} | \Vg{\theta}_{\mathrm{t}}, \Vg{\theta}_{\mathrm{d}}, \Vg{\theta}_x, \V{y})$, which was indirectly obtained above, will be necessary. Notice that, by Bayes' Theorem,
\begin{align}
    \begin{split}
        p(\V{x} |& \Vg{\theta}_{\mathrm{t}}, \Vg{\theta}_{\mathrm{d}}, \Vg{\theta}_x, \V{y}) \propto \\
        & p(\V{y} | \Vg{\theta}_{\mathrm{t}}, \Vg{\theta}_{\mathrm{d}}, \Vg{\theta}_x, \V{x})p(\V{x} | \Vg{\theta}_{\mathrm{t}}, \Vg{\theta}_{\mathrm{d}}, \Vg{\theta}_x) =\\
        & p(\V{y} | \Vg{\theta}_{\mathrm{t}}, \Vg{\theta}_{\mathrm{d}}, \Vg{\theta}_x, \V{x})p(\V{x} | n_0, M, \Vg{\theta}_x) = \\
        & [\delta(\V{y}_0 - \V{x}_0) \Ncal(\V{x}_1 | \V{y}_1, \sigma_{\mathrm{d}}^2\V{I}_M) \delta(\V{y}_2 - (\V{x}_2 + \V{v}_{\mathrm{t}}))] \times \\
	    &~~~~\Ncal(\V{K}\V{x}_0 + \V{U}_1\V{x}_1 + \V{U}_2\V{x}_2 | \V{0}, \sigma_e^2(\V{A}^T\V{A})^{-1}),
	\end{split}
\end{align}
as obtained in Equation~(\ref{eq:joint-dist-x-y}). This expression can be further simplified by noting that the second Gaussian depends essentially only on $\V{x}_1$, since its dependence on $\V{x}_0$ and $\V{x}_2$ is defined by the two Dirac's deltas. After using again the fact that a product of Gaussians is also Gaussian and calculating its mean and covariance matrix, one obtains
\begin{equation}
    \label{eq:post-dist-x}
    \begin{split}
	    \!\!p(\V{x} &| \Vg{\theta}_{\mathrm{t}}, \Vg{\theta}_{\mathrm{d}}, \Vg{\theta}_x, \V{y}) =\\
	    &\delta(\V{y}_0 - \V{x}_0) \Ncal(\V{x}_1 | \V{x}_1^{\text{MAP}}, \sigma_e^2\Vg{\Phi}^{-1}) \delta(\V{y}_2 - (\V{x}_2 + \V{v}_{\mathrm{t}})).
	\end{split}
\end{equation}

\section{Computation of the conditional distributions and sampling procedure}
\label{sec:cond-dist-sampling}
In this section, the conditional distributions which must be sampled from are explicitly computed, and the sampling procedure itself is detailed for each case. 

\subsection{Sampling $n_0$ and $M$} 
By using Bayes' Theorem and recalling the prior independence between $n_0$ and $M$, we have
\begin{align}
        \!\!\!\!p(n_0, M | \sigma_{\mathrm{d}}^2, \Vg{\theta}_{\mathrm{t}}, \V{x}, \Vg{\theta}_x, \V{y}) \!\propto\! p(\V{y} | \Vg{\theta}_{\mathrm{t}}, \Vg{\theta}_{\mathrm{d}}, \Vg{\theta}_x, \V{x})p(n_0)p(M),\!\!
\end{align}
where $p(\V{y} | \Vg{\theta}_{\mathrm{t}}, \Vg{\theta}_{\mathrm{d}}, \Vg{\theta}_x, \V{x})$ is given by Equation~(\ref{eq:likelihood-y}).
Note that this equation depends implicitly on $n_0$ and $M$, which influence the size of vectors $\V{y}_0$, $\V{y}_1$ and $\V{y}_2$. Therefore, this formula defines a complicated distribution that is not easy to sample from, requiring a Metropolis-Hastings step within the Gibbs sampler. 
The proposal distribution employed here is uniform over an interval whose length can be controlled by the user, centered at its respective last accepted values. Experiments indicated that small lengths are preferable, and a value of 10 is suggested.


\subsection{Sampling $\Vg{\theta}_{\mathrm{t}}$ and $\V{x}$} 

In order to sample from the joint posterior distribution of $\Vg{\theta}_{\mathrm{t}}$ and $\V{x}$ we recall the decomposition of $p(\Vg{\theta}_{\mathrm{t}}, \V{x} | \Vg{\theta}_{\mathrm{d}}, \Vg{\theta}_x, \V{y})$ in Equation~(\ref{eq:decomp-dist-thetat-x}). By using Bayes' Theorem, it implies that this joint sampling is performed by first sampling $\Vg{\theta}_{\mathrm{t}}$ from $p(\Vg{\theta}_{\mathrm{t}} | \Vg{\theta}_{\mathrm{d}}, \Vg{\theta}_x, \V{y}) \propto p(\V{y} | \Vg{\theta}_{\mathrm{t}}, \Vg{\theta}_{\mathrm{d}}, \Vg{\theta}_x)p(\Vg{\theta}_{\mathrm{t}})$ and then sampling $\V{x}$ from $p(\V{x} | \Vg{\theta}_{\mathrm{t}}, \Vg{\theta}_{\mathrm{d}}, \Vg{\theta}_x, \V{y})$.

The second step is quite straightforward, since this distribution was already computed and is given by Equation~(\ref{eq:post-dist-x}). Therefore, we only set $\V{x}_0 = \V{y}_0$ and $\V{x}_2 = \V{y}_2 - \V{v}_{\mathrm{t}}$, and sample $\V{x}_1$ from a Gaussian distribution with mean $\V{x}_1^{\text{MAP}}$ and covariance matrix $\sigma_{\mathrm{e}}^2\Vg{\Phi}^{-1}$. Note that $\V{x}_1$ is initialized simply with zeros, meaning that no previous
knowledge about the signal underlying the initial discontinuity is available.

The first step, sampling from $p(\Vg{\theta}_{\mathrm{t}} | \Vg{\theta}_{\mathrm{d}}, \Vg{\theta}_x, \V{y}) \propto p(\V{y} | \Vg{\theta}_{\mathrm{t}}, \Vg{\theta}_{\mathrm{d}}, \Vg{\theta}_x)p(\Vg{\theta}_{\mathrm{t}})$ is more complicated, and depends on whether the shape-based or the Gaussian Process model is being considered. The former is briefly recalled below for the sake of completeness, followed by an exposition of the latter.

\subsubsection{Shape-based model}
Note that the distribution $p(\Vg{\theta}_{\mathrm{t}}^{\mathrm{s}} | \Vg{\theta}_{\mathrm{d}}, \Vg{\theta}_x, \V{y}) \propto p(\V{y} | \Vg{\theta}_{\mathrm{t}}^{\mathrm{s}}, \Vg{\theta}_{\mathrm{d}}, \Vg{\theta}_x)p(\Vg{\theta}_{\mathrm{t}}^{\mathrm{s}})$, seen as a function of $\Vg{\theta}_{\mathrm{t}}^{\mathrm{s}}$, is very complicated, since $p(\V{y} | \Vg{\theta}_{\mathrm{t}}^{\mathrm{s}}, \Vg{\theta}_{\mathrm{d}}, \Vg{\theta}_x)$ is given by Equation~(\ref{eq:partial-likelihood-y}) and parameters in $\Vg{\theta}_{\mathrm{t}}^{\mathrm{s}}$ influence it through $\V{y}_2$ and the model in Equation~(\ref{eq:model-parametric-pulse}). Except for $V_{\mathrm{t}}$, whose posterior distribution can be verified to be Gaussian \cite{carvalho-avila-biscainho-aesbrasil-2014, carvalho-dsc}, another Metropolis-Hastings step within the Gibbs sampler must be employed to provide samples of the components of $\Vg{\theta}_{\mathrm{t}}^{\mathrm{s}}$. The chosen proposal distribution is multivariate Gaussian centered on the previous accepted value of the components of $\Vg{\theta}_{\mathrm{t}}^{\mathrm{s}}$ except $V_{\mathrm{t}}$. The diagonal covariance matrix of this proposal distribution is tuned by the user to keep the acceptance rate of this particular step around 50\%, as suggested by some authors to guarantee that the sample space is well explored in a reasonable computational time~\cite{givens-hoeting-computational-statistics}; particular values are recommended in Section \ref{sec:results}. This tuning can be made by increasing or decreasing the corresponding variances based on short runs of the simulated chain; despite being a classical procedure in Computational Statistics, this procedure was verified in experiments to be difficult to be performed, highly signal-dependent, and time-consuming, also possibly impacting in the convergence of the Gibbs sampler.

\subsubsection{Gaussian Process model}
Recall that in this scenario the only component of $\Vg{\theta}_{\mathrm{t}}^{\mathrm{G}}$ to be sampled within the Gibbs sampler is $\V{v}_{\mathrm{t}}^{\mathrm{G}}$, since $\sigma_f^2$ and $\sigma_{\ell}^2$ are kept constant. However, even if its conditional distribution is readily available, due to its high-dimensionality (in the order of thousands of time samples), sampling from it is computationally very expensive. Therefore, in this step we approximate a sample of this distribution by its mean, denoted by $\V{v}_{\mathrm{t}, \mathrm{mean}}^{\mathrm{G}}$, derived as follows.

By using Bayes' Theorem, we have that
\begin{align}
    \label{eq:post-thetat-g}
    \begin{split}
        p(\Vg{\theta}_{\mathrm{t}}^{\mathrm{G}} | \Vg{\theta}_{\mathrm{d}}, \Vg{\theta}_x, \V{y}) &= p(\V{v}_{\mathrm{t}}^{\mathrm{G}} | \Vg{\theta}_{\mathrm{d}}, \Vg{\theta}_x, \sigma_f^2, \sigma_{\ell}^2, \V{y}) \\
        &\propto p(\V{y} | \Vg{\theta}_{\mathrm{t}}^{\mathrm{G}}, \Vg{\theta}_{\mathrm{d}}, \Vg{\theta}_x)p(\V{v}_{\mathrm{t}}^{\mathrm{G}} | \sigma_f^2, \sigma_{\ell}^2),
    \end{split}
\end{align}
where $p(\V{y} | \Vg{\theta}_{\mathrm{t}}^{\mathrm{G}}, \Vg{\theta}_{\mathrm{d}}, \Vg{\theta}_x)$ has already been determined in Equations~(\ref{eq:partial-likelihood-y}) and~(\ref{eq:patial-likelihood-y-cont}), further simplified as Equations~(\ref{eq:partial-likelihood-y-simp-R}) and~(\ref{eq:partial-likelihood-y-simp-S}). 
By applying the Gaussian Process prior of Equation~(\ref{eq:gp-prior-tail}) in Equation~(\ref{eq:post-thetat-g}), the conditional posterior distribution for $\V{v}_{\mathrm{t}}^{\mathrm{G}}$ is given by
\begin{align}
    \begin{split}
        p(\V{v}_{\mathrm{t}}^{\mathrm{G}} &| \Vg{\theta}_{\mathrm{d}}, \Vg{\theta}_x, \sigma_f^2, \sigma_{\ell}^2, \V{y}) \propto \\ & \exp\left(-\frac{1}{2}(\V{v}_{\mathrm{t}}^{\mathrm{G}})^T\V{C}\V{v}_{\mathrm{t}}^{\mathrm{G}}\right)\exp\left(-\frac{1}{2}\V{z}^T\V{R}\V{z}\right)= \\
	    &\exp\left\{-\frac{1}{2}\left((\V{v}_{\mathrm{t}}^{\mathrm{G}})^T\V{C}\V{v}_{\mathrm{t}}^{\mathrm{G}} + \V{z}^T\V{R}\V{z}\right)\right\}.
    \end{split}
\end{align}
We must then compute the term $\V{z}^T\V{R}\V{z}$ in order to make explicit its dependence on $\V{v}_{\mathrm{t}}^{\mathrm{G}}$. To this end, note that it can be rewritten as:
\begin{align}
    \begin{split}
        \V{z}^T&\V{R}\V{z} = [\V{y}_0^T ~ (\V{y}_2 - \V{v}_{\mathrm{t}}^{\mathrm{G}})^T] \!\!
	    \left[
	    \begin{array}{c|c}
		    \V{R}_{11} & \V{R}_{12} \\
		    \hline
		    \V{R}_{21} & \V{R}_{22}
	    \end{array}
	    \right] \!\!\! \begin{bmatrix}\V{y}_0 \\ \V{y}_2 - \V{v}_{\mathrm{t}}^{\mathrm{G}} \end{bmatrix} \! = \\
	    & -\V{y}_0^T\V{R}_{12}\V{v}_{\mathrm{t}}^{\mathrm{G}} - (\V{v}_{\mathrm{t}}^{\mathrm{G}})^T\V{R}_{21}\V{y}_0 - \V{y}_2^T\V{R}_{22}\V{v}_{\mathrm{t}}^{\mathrm{G}} \\ &- (\V{v}_{\mathrm{t}}^{\mathrm{G}})^T\V{R}_{22}\V{y}_2 + (\V{v}_{\mathrm{t}}^{\mathrm{G}})^T\V{R}_{22}\V{v}_{\mathrm{t}}^{\mathrm{G}} \\ &+ \text{terms not depending on } \V{v}_{\mathrm{t}}^{\mathrm{G}}.
    \end{split}
\end{align}
Therefore, we have that
\begin{align}
    \begin{split}
	    (\V{v}_{\mathrm{t}}^{\mathrm{G}}&)^T\V{C}\V{v}_{\mathrm{t}}^{\mathrm{G}} +
	    \V{z}^T\V{R}\V{z} = \\ &-\V{y}_0^T\V{R}_{12}\V{v}_{\mathrm{t}}^{\mathrm{G}} - (\V{v}_{\mathrm{t}}^{\mathrm{G}})^T\V{R}_{21}\V{y}_0 - \V{y}_2^T\V{R}_{22}\V{v}_{\mathrm{t}}^{\mathrm{G}} \\
	    &- (\V{v}_{\mathrm{t}}^{\mathrm{G}})^T\V{R}_{22}\V{y}_2 + (\V{v}_{\mathrm{t}}^{\mathrm{G}})^T\V{R}_{22}\V{v}_{\mathrm{t}}^{\mathrm{G}} + (\V{v}_{\mathrm{t}}^{\mathrm{G}})^T\V{C}\V{v}_{\mathrm{t}}^{\mathrm{G}} \\
	    &+\text{terms not depending on } \V{v}_{\mathrm{t}}^{\mathrm{G}}.
	\end{split}
\end{align}

Since this expression is quadratic in $\V{v}_{\mathrm{t}}^{\mathrm{G}}$, the conditional posterior distribution for $\V{v}_{\mathrm{t}}^{\mathrm{G}}$ is a Gaussian whose mean vector and covariance matrix can be easily computed by completing the squares on the expression above, as indicated in the beginning of Section \ref{sec:computation-likelihood}. We then have that $p(\V{v}_{\mathrm{t}}^{\mathrm{G}} | \Vg{\theta}_{\mathrm{d}}, \Vg{\theta}_x, \sigma_f^2, \sigma_{\ell}^2, \V{y}) =  \Ncal(\V{v}_{\mathrm{t}, \mathrm{mean}}^{\mathrm{G}}, \Vg{\Sigma})$, where
\begin{align}
    \begin{split}
        \V{v}_{\mathrm{t}, \mathrm{mean}}^{\mathrm{G}} = [\V{R}_{22} + \V{R}_{22}^T + \V{C}^{-1} + \V{C}^{-T}]^{-1} \times \\ [(\V{R}_{12}^T + \V{R}_{21})\V{y}_0 + (\V{R}_{22}^T + \V{R}_{22})\V{y}_2]
    \end{split}
\end{align}
\begin{align}
    \Vg{\Sigma} &= \left[\frac{1}{2}(\V{R}_{22} + \V{R}_{22}^T + \V{C}^{-1} +  \V{C}^{-T})\right]^{-1}.
\end{align}

Finally, note that if we have two overlapping pulses, their respective tails are being modeled by Gaussian Processes, and when the respective tails overlap, the posterior distribution of the superimposed pulse will still be Gaussian. Therefore, this modeling may allow a simpler treatment of this heretofore complicated scenario. However, preliminary tests indicated that the estimation of the AR model parameters in order to interpolate the signal underlying an initial discontinuity superimposed to an unfinished pulse tail is problematic; this issue is left to be addressed in a future work.


\subsection{Sampling $\sigma_{\mathrm{d}}^2$}
This is the last step of the Gibbs sampler. To compute the required posterior distribution, we use Bayes' Theorem together with the prior independence once again to obtain
\begin{align}
	p(\sigma_{\mathrm{d}}^2 | n_0, M, \Vg{\theta}_{\mathrm{t}}, \V{x}, \Vg{\theta}_x, \V{y}) \propto p(\V{y} | \Vg{\theta}_{\mathrm{t}}, \Vg{\theta}_{\mathrm{d}}, \Vg{\theta}_x, \V{x})p(\sigma_{\mathrm{d}}^2).
\end{align}
Now, the dependence of $p(\V{y} | \Vg{\theta}_{\mathrm{t}}, \Vg{\theta}_{\mathrm{d}}, \Vg{\theta}_x, \V{x})$ on $\sigma_{\mathrm{d}}^2$ is very simple: as can be seen in Equation~(\ref{eq:likelihood-y}), it is just a scale parameter for the distribution. Therefore, the Inverse Gamma prior for $\sigma_{\mathrm{d}}^2$ implies that $p(\sigma_{\mathrm{d}}^2 | n_0, M, \Vg{\theta}_{\mathrm{t}}, \V{x}, \Vg{\theta}_x, \V{y})$ is also an Inverse Gamma, with parameters given by
\begin{align}
	\alpha = \alpha_{\mathrm{d}} + \frac{M}{2}
\end{align}
and
\begin{align}
	\beta = \beta_{\mathrm{d}} + \frac{1}{2}\sum\limits_{i = 0}^{M-1}\V{v}_{\mathrm{d}}(n_0 + i)^2.
\end{align}

\section{Results}
\label{sec:results}
The performance of the proposed algorithm was evaluated through tests performed in three distinct scenarios, namely:
\begin{itemize}
    \item [(A) ] A real signal artificially distorted by a pulse following Equation~\ref{eq:model-parametric-pulse}, in order to assess the method's accuracy and convergence, by means of a complete statistical analysis of the simulated chains;
    
    
    \item [(B) ] Three real signals artificially degraded by pulses following Equation~\ref{eq:model-parametric-pulse}, whose restoration results are compared with other methods from the literature;
    
    \item [(C) ] Two real degraded signals, informing us about the method's capability of dealing with real distortions.
\end{itemize}
The tests were run in a PC with a quadcore processor operating at 1.60 GHz clock and 8 GB of RAM. All signals are monophonic, sampled at 44.1 kHz with 16-bit precision, and implementations are in MATLAB\texttrademark \footnote{The MathWorks, Inc., \texttt{http://www.mathworks.com/}.}.

\subsection{Statistical analysis of the simulated chains}
This test was performed over the first 3 s of a 13-s excerpt from a musical track, which consists of orchestral music containing a slowly varying string passage with some percussion in the last 5 seconds. The signal was degraded by a single pulse following Equation~\ref{eq:model-parametric-pulse}. The main goal of this test was to simulate long runs of the proposed algorithm, considering both the shape-based and Gaussian Process modeling for the pulse tail, and evaluate its convergence properties. 
Table~\ref{tab:exp-1} summarizes the results obtained, which will be discussed along this Section. Its first nine lines refers to the shape-based model parameters, and the last three ones to the Gaussian Process counterparts. The parameters common to both models are distinguished by the superscripts ``s"~and ``G", respectively.

\begin{table*}
\centering
\caption{Summary of statistical properties of the chains simulated in experiment (A). }
\label{tab:exp-1}
\begin{tabular}{@{}lccccccc@{}}
\toprule
Parameter               & True value & Initial value & Credible interval at 95\% & Estimate I & Estimate II & Variance of proposal distribution & Acceptance rate \\ \midrule
$n_0^{\mathrm{s}}$                   & 497 & 500           & $[497; ~ 497]$ & 497 & 497 & 10 & 0.0096 \\
$M^{\mathrm{s}}$                     & 10 & 16            & $[10; ~ 10]$& 10 & 10 & 10 & 0.0096 \\
$\sigma_{\mathrm{d}}^{2^{\mathrm{s}}}$ & 0.5 & 0.4126        & $[0.3036; ~ 2.1272]$ & 0.8607 & 0.8600 & -- & -- \\
$V_{\mathrm{t}}$        & 0.3 & 0.1           & $[0.2628; ~ 0.3542]$   & 0.3122         & 0.3100 & -- & -- \\
$\tau_m$                & 0.07 & 0.1           & $[0.0548; ~ 0.0890]$   & 0.0680          & 0.0683 & $1.5 \times 10^{-5}$ & 0.4457 \\
$\tau_f$                & 0.013 & 0.19          & $[0.0066; ~ 0.0180]$   & 0.0114          & 0.0138 & $5 \times 10^{-7}$ & 0.3407 \\
$\fmax$                 & 60 & 50            & $[34.1076; ~ 77.3995]$ & 54.4956       & 46.4982 & $6$ & 0.7472 \\ 
$\fmin$                 & 20 & 30            & $[18.2054; ~ 20.1479]$ & 19.3426       & 19.3351  & $0.6$ & 0.2887 \\
$\phi$                  & 0 & 0.5           & $[0.0196; ~ 0.9382]$   & 0.4019         & 0.3668  & $10^{-2}$ & 0.5130 \\ \midrule
$n_0^{\mathrm{G}}$                   & 497 & 500           & $[497; ~ 497]$ & 497 & 497 & 10 & 0.0180 \\
$M^{\mathrm{G}}$                     & 10 & 16            & $[10; ~ 10]$& 10 & 10 & 10 & 0.0180 \\
$\sigma_{\mathrm{d}}^{2^{\mathrm{G}}}$ & 0.5 & 0.4126        & $[0.3510; ~ 2.3905]$ & 0.9586 & 0.8812 & -- & -- \\ \bottomrule
\end{tabular}
\end{table*}

\begin{figure}
    \centering
    \includegraphics[width=8.5cm]{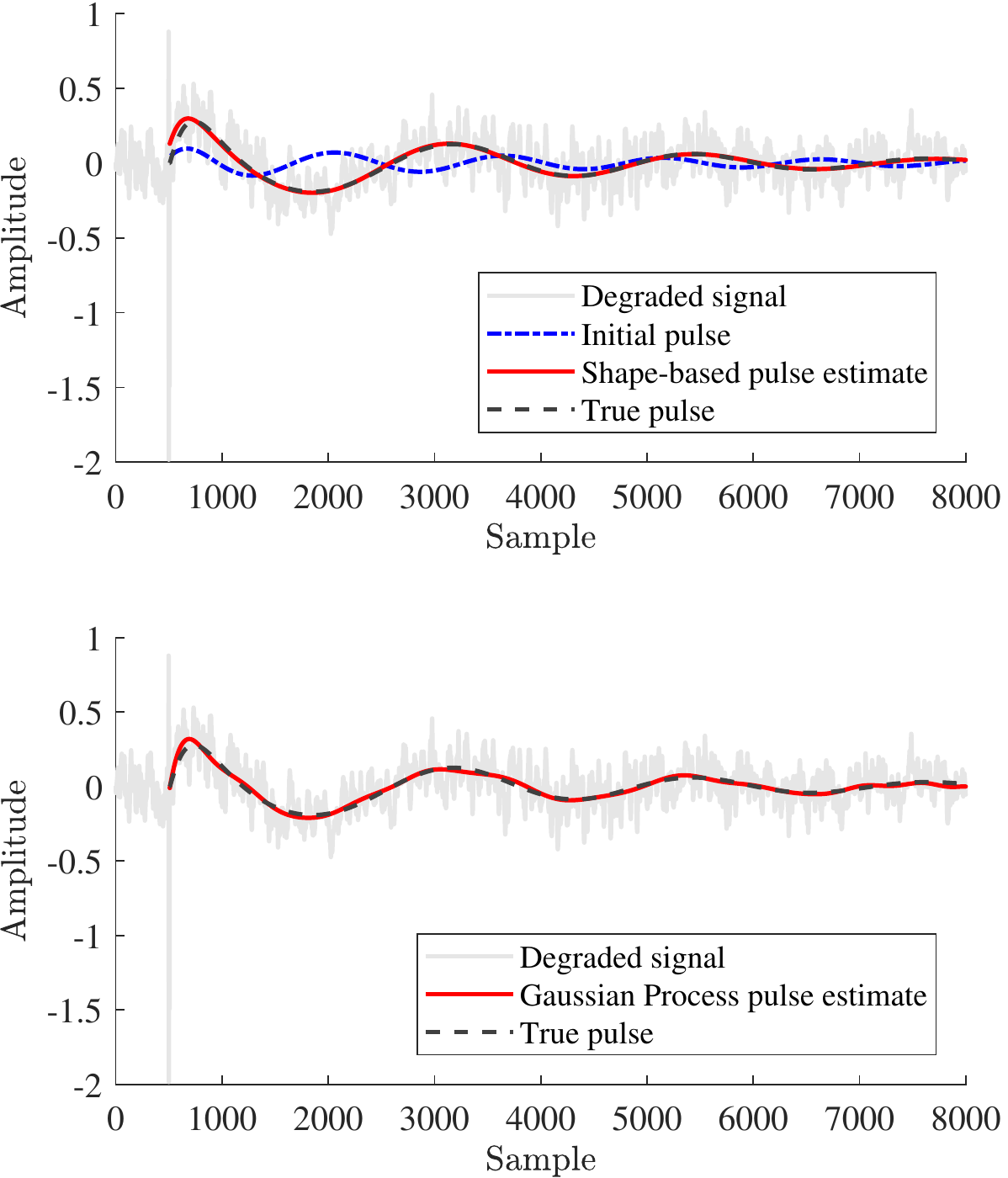}
    \caption{Comparison between true and estimated pulses in experiment (A): light continuous and dark dashed lines represent the underlying distorted signal and the superimposed pulse, respectively, in both graphs; in the upper panel, the solid dark continuous line is the pulse estimate by the shape-based algorithm, and the point-dashed line is the pulse generated by the initial values of $\Vg{\theta}_{\mathrm{t}}^{\mathrm{s}}$; in the lower panel, the solid dark line is the pulse estimated by the Gaussian Process algorithm.}
    \label{fig:exp_1}
\end{figure}

The 3-s degraded signal was given as input to the initialization procedure with parameters $L = 16$, $\xi = 0.3$, $c = 5$, and $f_{\mathrm{co}} = 3$ kHz. Once the single pulse had been localized through the initialization procedure, a segment of $N = 8,\!000$ samples containing the pulse was processed. Within this block, the initial estimate of the pulse start was indexed as $n_0=500$ (its true value is 3 samples earlier), and the parameters of the order-40 AR model used to infer the underlying signal in the initial discontinuity were estimated from the first 450 samples. The initial estimates for $M$ and $\sigma_{\mathrm{d}}^2$ are displayed in the third column of Table~\ref{tab:exp-1}, which also contains the initial values for $\Vg{\theta}_{\mathrm{t}}^{\mathrm{s}}$: despite seeming numerically close to their respective true values (displayed in the second column of Table~\ref{tab:exp-1}), they model quite distinct pulses, as illustrated in Figure \ref{fig:exp_1}. 


A single chain with $10,\!000$ iterations was simulated for the proposed algorithm with the {\bf shape-based model}, being the first $5,\!000$ discarded as burn-in time. Each iteration lasted for approximately $1.6$ s. Since the autocorrelation plots of the sampled values after the burn-in time showed a highly correlated chain, of every 50 samples only one was considered and the rest were discarded, which resulted in 100 approximately uncorrelated samples for the posterior distribution of $\V{x}$, $\Vg{\theta}_{\mathrm{d}}$, and $\Vg{\theta}_{\mathrm{t}}^{\mathrm{s}}$. Recall that estimates are obtained by averaging these sampled values, as mentioned in Section \ref{sec:description-algorithm}.

The estimates for $\Vg{\theta}_{\mathrm{d}}$ and $\Vg{\theta}_{\mathrm{t}}^{\mathrm{s}}$ are displayed in the column named ``Estimate I"~of Table~\ref{tab:exp-1}, whose left column contains the respective estimated credible intervals at 95\%. Note that, except for $\phi$, all credible intervals contain the respective true values, indicating that the algorithm was capable of correctly identifying most of the degrading parameters. It is also important to remark that after the burn-in time all proposed values for $n_0$ and $M$ distinct from the actual ones were rejected, meaning that their respective posterior distributions are highly concentrated around these values. Such behavior is expected in this scenario, since the artificial degradation was introduced in a signal without other defects, such as broadband additive noise or clicks, which could increase the variances of the distributions.

The penultimate column of Table~\ref{tab:exp-1} displays the variance of the proposal distribution for the components of $\Vg{\theta}_{\mathrm{d}}$ and $\Vg{\theta}_{\mathrm{t}}^{\mathrm{s}}$, except for $\sigma_{\mathrm{d}}^2$ and $V_{\mathrm{t}}$, which have closed form and well-known full conditional distributions. For $n_0$ and $M$, the value corresponds to the aforementioned length-10 discrete uniform proposal. The values for the other variables were manually tuned by simulating shorter chains, and yield the acceptance rates displayed in the last column of Table~\ref{tab:exp-1}. Note that these rates are in a broad range around the desired 50\%, a behavior also observed when using the same variances with other signals in experiments (B) and (C). In fact, the choice of such values allowed satisfactory restored signals to be obtained in all scenarios tested without the need to fine-tune the algorithm separately for each situation.

Finally, in an attempt to circumvent the computational burden of the algorithm, another estimate was performed by considering only $1,\!000$ iterations of the chain: the first $500$ samples were discarded as burn-in phase, and the last $500$ were averaged without discarding any sample, thus not masking the chain correlation. These estimates are displayed in the column ``Estimate II"~of Table~\ref{tab:exp-1}; it can be noticed that they are also close to their respective true values and within the respective credible intervals at 95\%, except for $\phi$. This last procedure will be preferred over the previous in the next tests with the shape-based model, since it was found that it produces equally satisfactory results with less computational effort.

\begin{figure}
    \centering
    \includegraphics[width=8.5cm]{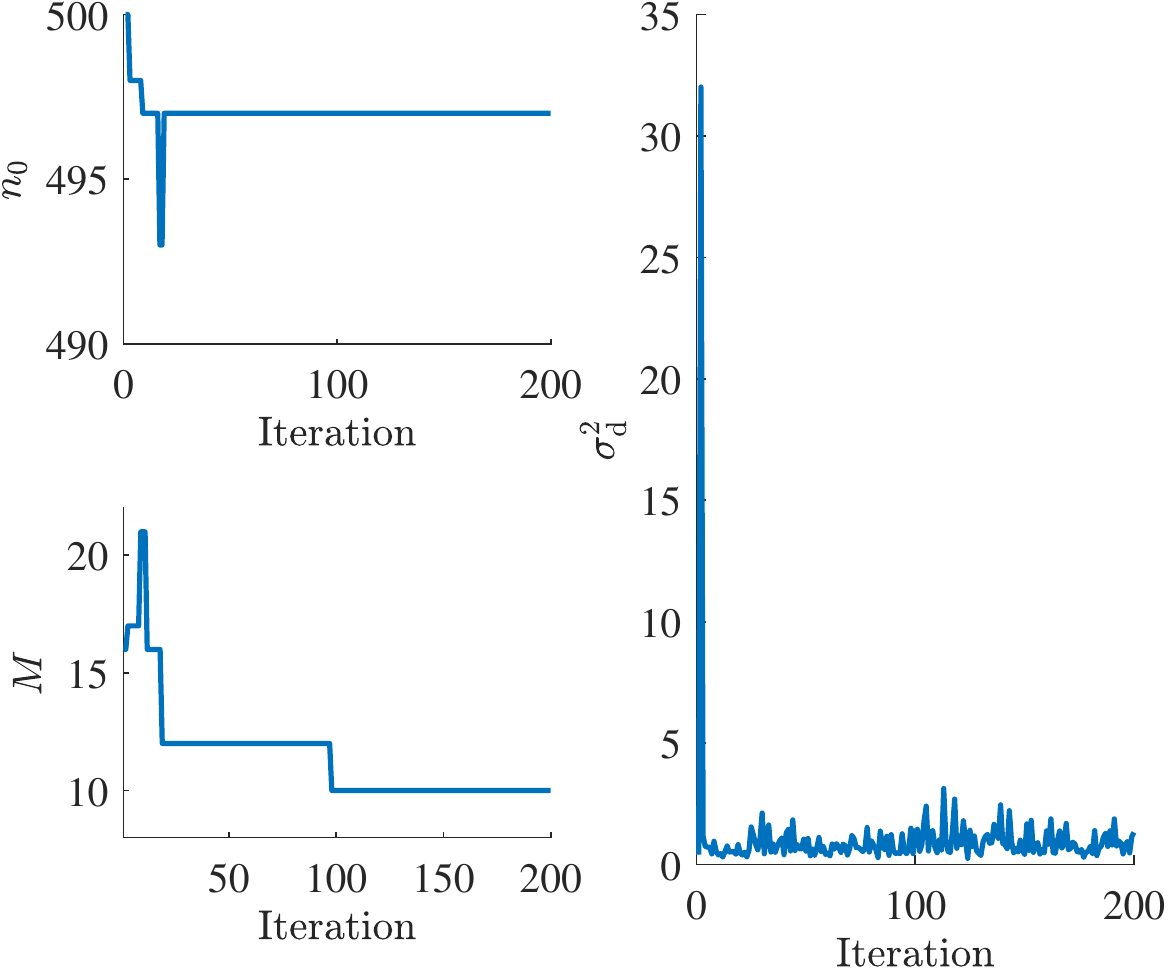}
    \caption{First 200 iterations of the simulated chain for $n_0$, $M$, and $\sigma_{\mathrm{d}}^2$ in experiment (A) for the Gaussian Process model. The rest of the iterations show similar behavior: $n_0$ and $M$ are constant, and analogous oscillations occur for $\sigma_{\mathrm{d}}^2$.}
    \label{fig:exp_1_GP}
\end{figure}

As for the {\bf Gaussian Process model}, a chain of $1,\!000$ iterations was simulated, each iteration lasting for approximately 9~s, with the same initialization procedure described before. Note that in addition to the initialization procedure parameters, the only parameters to be tuned in this algorithm are the supports of the proposal distributions for $n_0$ and $M$, both set to 10. The behavior of the simulated chain for these variables, as well as for $\sigma_{\mathrm{d}}^2$, is similar to that observed in the shape-based case, and the first 200 iterations are displayed in Figure~\ref{fig:exp_1_GP}. It can be seen that the sampled values of $n_0$ and $M$ rapidly converge to their respective true values, and contrarily to the shape-based case, it was verified that the simulated chain for $\sigma_{\mathrm{d}}^2$ does not exhibit the autocorrelation problem, dispensing with the thinning procedure.
A summary of the simulation is displayed in the last three lines of Table~\ref{tab:exp-1}: the estimates obtained by averaging the samples after the burn-in time of $500$ iterations is displayed in the column named ``Estimate I", and the estimated credible intervals at 95\% are in its left column. In order to reduce the computational impact of this model, it would be enough to simulate a chain with 200 iterations and discard the first 150 as burn-in time, leading to the estimates in the ``Estimate II"~column of Table~\ref{tab:exp-1}. This procedure will be adopted in the next tests with the Gaussian Process model.

The true pulse and the pulses estimated by both the shape-based and Gaussian Process models are displayed in Figure~\ref{fig:exp_1}. In the upper panel it can be seen that, despite the inaccuracy in the estimate of $\phi$ in the shape-based case, both true and estimated pulses are visually similar, indicating the capability of the algorithm to suitably identify the degradation; the pulse generated by the initial value for $\Vg{\theta}_{\mathrm{t}}^{\mathrm{s}}$ is also displayed. As shown in the lower panel, the Gaussian Process algorithm is also capable of satisfactorily estimating the pulse, despite the stationary squared-exponential covariance kernel employed. 

Despite not being the case illustrated in Figure~\ref{fig:exp_1}, due to the stationary nature of the squared-exponential covariance kernel employed in the Gaussian Process model, the estimated pulse can be distant from zero at the end of the processed block. In this situation, the restored excerpt, when replaced on the original signal, will exhibit a discontinuity that may generate audible artifacts. This issue can be solved by a simple heuristics: fading-out the last samples of the estimated pulse linearly to zero before being subtracted from the degraded signal. For a block of the size considered in experiments (A) and (B), fading-out the last $1,\!000$ samples showed good results; this procedure was not necessary in experiment (C). 

In short, these experiments illustrate the following: 1) the initialization procedure is capable of providing accurate estimates of the pulse location, which are promptly corrected by the algorithm under both shape-based and Gaussian Process modeling; 2) when the variances of the proposal distributions for the Metropolis-Hastings steps in the shape-based case are properly tuned and sufficiently long chains are simulated, the algorithm is capable of correctly identifying the pulse parameters; 3) despite the inaccuracy of the squared-exponential covariance kernel employed on the Gaussian Process model, it is also capable of identifying the pulse.




\subsection{Comparison with previously proposed methods}
Three signals were evaluated in this test, with both the shape-based and Gaussian Process models: the same 13-second long excerpt of orchestral music mentioned in Experiment (A), an 8-second long excerpt of jazz quartet music with drums, bass, guitar and clarinet, and an 11-second long excerpt of Brazilian music with singing male voice and percussion. These signals are referred to as ``classical", ``jazz", and ``Brazilian", and were artificially corrupted with 17, 11, and 14 non-overlapping and uniformly spaced pulses, respectively, following Equation~\ref{eq:model-parametric-pulse}.

\begin{figure}
    \centering
    \includegraphics[width=8.5cm]{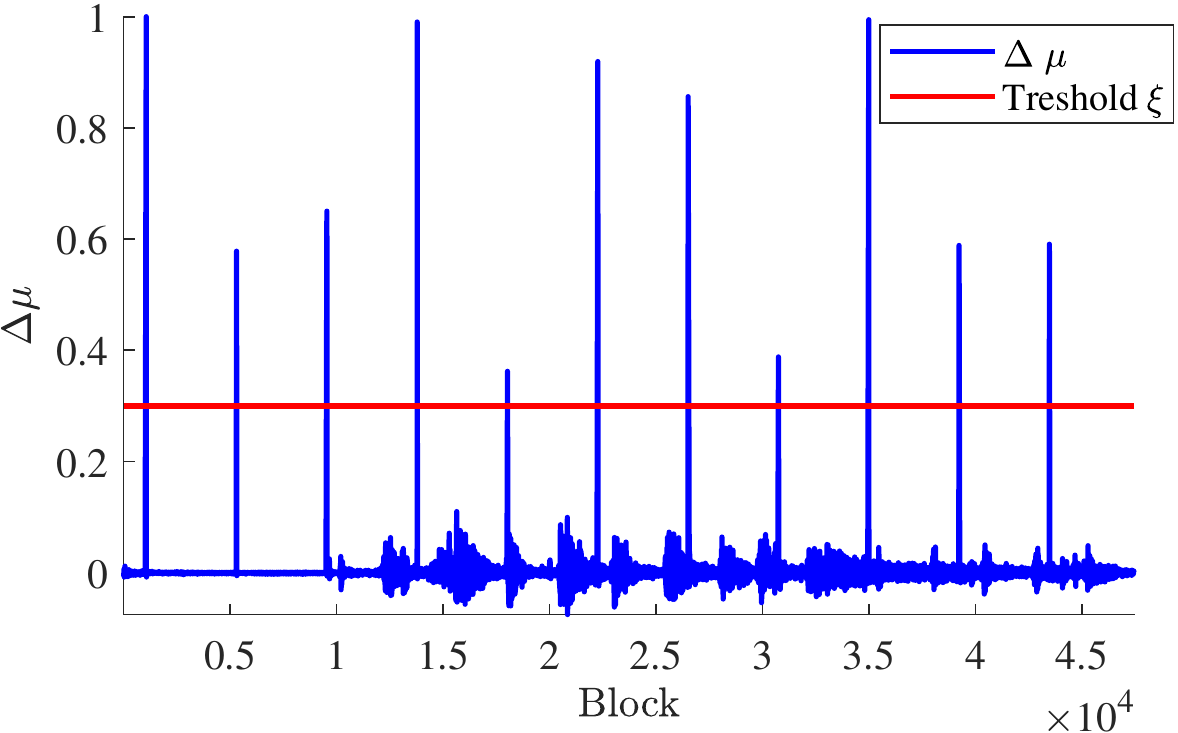}
    \caption{Function $\Delta \mu$ -- output of the initialization procedure (with associated threshold) for the ``jazz"~signal in experiment (B).}
    \label{fig:delta-mu}
\end{figure}

All signals were given as input to the initialization procedure with the same setup as before. For illustration, its output $\Delta \mu$ is shown for the ``jazz"~signal in Figure~\ref{fig:delta-mu}. Note that all pulses are identified; the same was verified in the other two test signals. An excerpt of size $N = 8,\!000$ containing each pulse was separately processed following the shorter procedures previously described, namely: for the {\bf shape-based model}, $1,\!000$ iterations with burn-in of $500$ for each pulse, $\Vg{\theta}_{\mathrm{t}}^{\mathrm{s}}$ initialized as in the third column of Table~\ref{tab:exp-1}, and variances of proposal distributions (except for $V_{\mathrm{t}}$) as in the penultimate column of Table~\ref{tab:exp-1}; for the {\bf Gaussian Process model}, $200$ iterations with burn-in of $150$. The estimates are obtained by averaging the samples obtained after the burn-in time, and the initial estimates for $n_0$, $M$, and $\sigma_{\mathrm{d}}^2$ are the outputs of the initialization procedure.

In order to objectively assess the quality of the restored signal, we evaluated it through the PEAQ (Perceptual Evaluation of Audio Quality)~\cite{itu, kabal-peaq} algorithm, tailored to compare the perceived quality of a wideband audio signal against a reference signal after mapping them to a perceptual domain that models the operation of the human auditory system. The output score is a number from 0 (imperceptible difference from the reference signal) to -4 (very annoying difference). Table~\ref{tab:PEAQ} compares the PEAQ scores received by the proposed method with both modeling possibilities with those of~\cite{esquef-et-al-long-pulse} (based on the TPSW filtering), \cite{esquef-welter} (based on the EMD),
and~\cite{godsill-tan-long-pulse} (based on the AR separation technique). The original signals and their restored counterparts using these methods are available from \cite{esquef-web}. As required by the PEAQ algorithm, signals were previously upsampled to 48 kHz. Note that two instances of the ``classical"~signal were considered: its entirety and its first 8 s, which excludes the mainly percussive part.
\begin{table}
\centering
\caption{PEAQ grades for the artificially degraded signals in experiment (B) and its respective restored versions. The closer the grade is to zero, the cleaner the signal.}
\label{tab:PEAQ}
\begin{tabular}{@{}lcccc@{}}
\toprule
Signal             & Jazz & Classical & \makecell{Classical \\ (no percussion)} & Brazilian \\ \midrule
Distorted          & -1.7427   & -1.6874 & -1.5398 & -1.9499 \\
Shape-based        & -0.5561   & -0.7108 & -0.3276 & -0.7814 \\
Gaussian Process   & -0.3627   & -0.2902 & -0.1266 & -0.2341 \\
TPSW               & -0.4499   & -0.3140 & -0.2640 & -0.2606 \\
EMD                 & -0.2893   & -0.2245 & -0.2081 & -0.2230 \\
AR                 & -0.5484   & -0.1947 & -0.1534 & -0.1489 \\ \bottomrule
\end{tabular}
\end{table}

Note that the proposed algorithm with both modeling possibilities returns signals with substantially improved PEAQ grades than its corrupted counterparts, being the performance of the Gaussian Process model always better than the shape-based one. Both models show their better results in the initial excerpt of the ``classical"~signal, a behavior explained by two aspects: 1) in this case the underlying signal easily fulfills the hypothesis of being similar to white noise in the time-scale of the tail, as can be seen in Figure~\ref{fig:exp_1}; and 2) the tuning parameters of the shape-based algorithm were the same as in experiment (A), being then tailored to a similar underlying signal. For the sake of completeness, the PEAQ scores of the degraded and restored (by both the shape-based and Gaussian Process algorithms) signals in experiment (A) were -2.6513, -0.1873, and -0.1471, respectively, indicating that the shape-based model is capable of producing results similar to the Gaussian Process model, when properly tuned.

Despite not providing much information from the perceptual viewpoint, the signal-to-noise ratio (SNR) measured in dB for the same set of signals is presented in Table~\ref{tab:SNR}. It can be seen that the proposed method always return restored signals with improved SNR, but this improvement is substantially lower for signals with substantial percussive content. Let us expand this discussion by considering the ``Brazilian"~signal as an example. Informal listening tests indicates that its restoration when using the Gaussian Process model sounds identical to the non-degraded signal, whereas when employing the shape-based model several remnants of the initial discontinuities are noticeable. The respective PEAQ grades reflects this impression, but their SNR indicates the opposite.
\begin{table}
\centering
\caption{SNR (in $\mathrm{dB}$) for the artificially degraded signals in experiment (B) and its respective restored versions. The higher the value, the closer to the cleaner signal.}
\label{tab:SNR}
\begin{tabular}{@{}lcccc@{}}
\toprule
Signal             & Jazz & Classical & \makecell{Classical \\ (no percussion)} & Brazilian \\ \midrule
Distorted          & 3.9864   & 4.1028 & 4.2073 & 5.1588 \\
Shape-based        & 14.5028   & 13.9956 & 18.8943 & 13.3625 \\
Gaussian Process   & 10.8130   & 15.7397 & 22.3557 & 7.9281 \\
TPSW               & 17.2948   & 18.8036 & 21.3458 & 11.8810 \\
EMD           & 18.6073  & 20.6855 & 24.3433 & 13.5084 \\
AR                 & 17.2558   & 18.9781 & 22.2038 & 9.1398 \\ \bottomrule
\end{tabular}
\end{table}

\begin{figure}
    \centering
    \includegraphics[width=8.5cm]{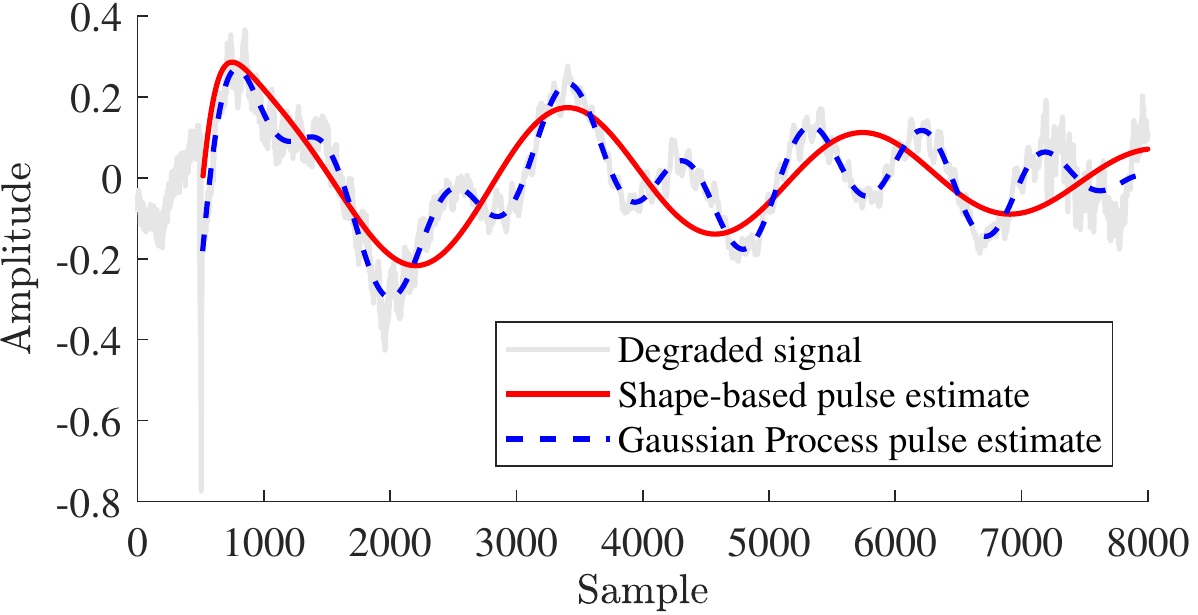}
    \caption{Excerpt processed containing the fifth pulse of the ``Brazilian"~signal (clear solid line), and pulse estimates using the shape-based (solid dark line) and Gaussian Process (dashed dark line).}
    \label{fig:exp_3_BR}
\end{figure}

To illustrate this case, in Figure~\ref{fig:exp_3_BR} the input excerpt containing the fifth pulse of the degraded version of the ``Brazilian"~signal is displayed, as well as the estimated pulse from both models. It can be seen that the Gaussian Process estimate also captures fluctuations intrinsic to the underlying signal that are not generated by the pulse, implying the observed low SNR. This can be seen as a disadvantage of this model when compared to the shape-based one. However, the latter has not showed itself capable of correctly interpolating the signal underlying the initial discontinuity, whereas the former leaves no audible remnant of the degradation, being the only exception noticed on the percussive part of the ``classical"~signal.

In summary, the proposed algorithm with the Gaussian Process model yields a signal with marginally better or similar PEAQ score when compared to its competitors. In practice, these numbers mean that the restored signals sound almost identical, a fact that was verified by informal listening tests. However, our proposed algorithm automatically locates the initial discontinuity and its duration, improves these initial estimates and interpolates the underlying signal; in addition, it produces a very accurate estimate of the pulse tail. Even so, it requires significantly less user intervention than the shape-based model.

\subsection{Real degraded signals}
The two signals considered in this scenario are: an excerpt of 3 seconds from cylinder recording number 154 from~\cite{catalan-cylinders}, approximately from 1907, containing a severe pulse at the beginning when the music that will be played is introduced; and an excerpt of 11 seconds from a piece by Chopin for solo piano, containing six pulses. Both are sampled at 44.1 kHz with 16-bit resolution. Parameters for the initialization procedure for each signal are shown in Table~\ref{tab:param-cyl-chop}. They were manually tuned in order to correctly identify the pulses only, and not other defects also present within the signal, such as clicks and crackles. Essentially, signals with more background noise require a higher value of $f_{\mathrm{co}}$ and $c$, but since the initialization procedure is very fast, lasting less than 1 s, some trial and error is not tedious or time-consuming.
\begin{table}
\centering
\caption{Parameters of the initialization procedure for the ``cylinder"~and ``Chopin"~signals in experiment (C).}
\label{tab:param-cyl-chop}
\begin{tabular}{@{}ccc@{}}
\toprule
                  & \multicolumn{2}{c}{Signal}                       \\
Parameter         & Cylinder                & Chopin                 \\ \midrule
$L$               & 64                      & 64                     \\
$\xi$             & 0.4                     & 0.3                    \\
$c$               & 31                      & 7                      \\
$f_{\mathrm{co}}$ & 10,000 & 3,000 \\ \bottomrule
\end{tabular}
\end{table}

In both cases, since the tail of the pulse is not too long, an excerpt of only 2,000 samples containing the degradation was processed. When considering the \textbf{Gaussian Process model}, for each pulse the Gibbs sampler ran for 500 iterations (lasting $\approx0.5$ s each), of which the first 400 were discarded; regarding the \textbf{shape-based model}, for each pulse $5,\!000$ iterations (lasting $\approx0.2$ s each) were run, of which the first $2,\!500$ were discarded. The initial value of $\theta_{\mathrm{t}}^{\mathrm{s}}$ was the same as in the third column of Table~\ref{tab:exp-1}.

\begin{figure}
    \centering
    \includegraphics[width=8.5cm]{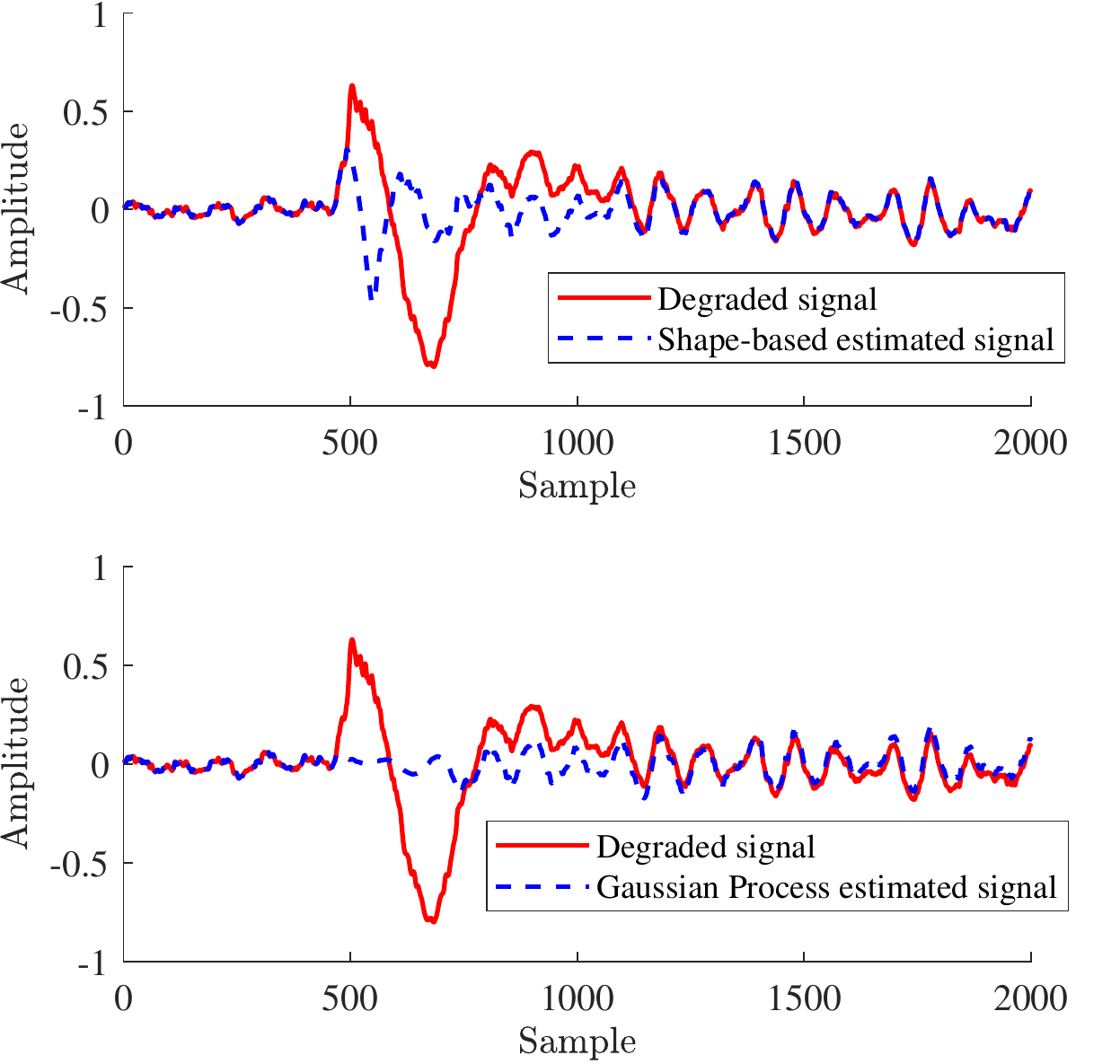}
    \caption{Processed excerpt of ``Cylinder"~signal (solid line) and its restored versions (dashed lines): shape-based in the upper panel and Gaussian Process in the lower one.}
    \label{fig:cil}
\end{figure}

Since in this scenario we do not have a reference signal, it is not possible to compute the PEAQ grade, so we will present a brief discussion about the restoration on both cases:
\begin{itemize}
    \item In the cylinder excerpt, the restoration seems almost perfect with the Gaussian Process model. A very light click remains where the pulse originally started, although almost masked by the high background noise present throughout the signal. On the other hand, there is no vestige of the low-frequency content of the pulse tail. The shape-based algorithm was also capable of removing the pulse tail and attenuate the initial discontinuity, but its remnant is much more perceptible. The processed excerpt and the restored signals can be seen in Figure~\ref{fig:cil}.
    
    \item In the Chopin excerpt, the restoration in both modeling scenarios sounded almost identical: the remnant of the initial discontinuities, this time in the form of a short-duration and low-level hiss, is more apparent because of the lower level of background noise in the signal. However, once more there is also no perceptible vestige of the low-frequency content of the pulse in the restored signal.
\end{itemize}

This small residue left after the interpolation of the signal underlying the initial discontinuity is much probably due to the impairment of the AR modeling by the presence of background noise, as discussed in~\cite{godsill-rayner}. This issue could be worked around by including the estimation of its parameters in the Gibbs sampler instead of keeping them constant, or by employing a more sophisticated interpolation scheme. However, the high impact on the overall computational load found in preliminary tests suggested that the simplified approach should be adopted and that this point should be further investigated with a view to a new version of the algorithm. However, it should be stressed that, as in the controlled experiments, the proposed algorithm was shown to be capable of identifying the location and extent of the initial discontinuity as well as the tail of the pulse when dealing with real degraded signals, a much more challenging scenario where defects other than the long pulses may also be present.



\section{Conclusion}
\label{sec:conclusion}
In this paper we presented a novel algorithm for the restoration of audio signals degraded by long pulses with significant low-frequency content that circumvents two main drawbacks of previous works: a large quantity of hyper-parameters difficult to interpret that must be adjusted manually, and the need to know the location of the pulse beforehand. These issues were addressed through modeling of the pulse tail via Gaussian Process and a Bayesian framework that includes location as a random quantity to be estimated. The signal underlying the initial discontinuity is also interpolated along the procedure. In order to accelerate the convergence of the Gibbs sampler, we employed an efficient initialization algorithm based on the time-frequency content of the degraded signal that roughly locate the beginning of the degradation. In controlled experiments, when compared to previous approaches our algorithm shows a slightly better or similar performance, according to the PEAQ grades. These results are confirmed by the experiments with real degraded signals.

Some improvements to the proposed algorithm will be addressed in future works: 1) make it able to handle overlapping pulses; and 2) refine the restored signal interpolation in the region of the initial discontinuity, which is likely to be impaired by other degradations such as background noise. Both points are linked to the difficulty of estimating the AR model parameters and further interpolate the restored signal in the region of the discontinuities in noisy scenarios; this is the next research target.

Another possible improvement is to sub-sample the pulse tail before the Gibbs sampler. This will significantly decrease the computational cost of the algorithm, since at each iteration a matrix of size approximately $N \times N$ must be inverted. The sub-sampling can be justified by the fact that it implies no relevant loss of information on the tail, since it usually contains much lower frequencies than the typical underlying signal. Finally, from the statistical viewpoint the adoption of Variational Inference instead of MCMC in the estimation procedure should be addressed in a future work aiming at an acceleration of the entire procedure.


%



\section*{Acknowledgment}
The authors would like to thank Paulo Antonio Andrade Esquef for providing the signals used in the experiment with artificially degraded signals, and the ``Chopin"~signal employed on the real degraded scenario.


\ifCLASSOPTIONcaptionsoff
  \newpage
\fi



\bibliographystyle{IEEEtran}
\bibliography{bare_jrnl}

\begin{thebibliography}{10}
\providecommand{\url}[1]{#1}
\csname url@samestyle\endcsname
\providecommand{\newblock}{\relax}
\providecommand{\bibinfo}[2]{#2}
\providecommand{\BIBentrySTDinterwordspacing}{\spaceskip=0pt\relax}
\providecommand{\BIBentryALTinterwordstretchfactor}{4}
\providecommand{\BIBentryALTinterwordspacing}{\spaceskip=\fontdimen2\font plus
\BIBentryALTinterwordstretchfactor\fontdimen3\font minus
  \fontdimen4\font\relax}
\providecommand{\BIBforeignlanguage}[2]{{%
\expandafter\ifx\csname l@#1\endcsname\relax
\typeout{** WARNING: IEEEtran.bst: No hyphenation pattern has been}%
\typeout{** loaded for the language `#1'. Using the pattern for}%
\typeout{** the default language instead.}%
\else
\language=\csname l@#1\endcsname
\fi
#2}}
\providecommand{\BIBdecl}{\relax}
\BIBdecl

\bibitem{vaseghi-phd}
S.~V. Vaseghi, ``Algorithms for restoration of archived gramophone
  recordings,'' \uppercase{P}h.\uppercase{D}. Thesis, University of Cambridge,
  Cambridge, UK, 1988.

\bibitem{vaseghi-frayling-long-pulse}
S.~V. Vaseghi and R.~Frayling-Cork, ``Restoration of old gramophone
  recordings,'' \emph{Journal of the Audio Engineering Society}, vol.~40,
  no.~10, pp. 791--801, 1991.

\bibitem{godsill-rayner}
S.~J. Godsill and P.~J.~W. Rayner, \emph{Digital Audio Restoration - A
  Statistical Model Based Approach}.\hskip 1em plus 0.5em minus 0.4em\relax
  Cambridge, UK: Cambridge University Press, 1998.

\bibitem{godsill-tan-long-pulse}
S.~J. Godsill and C.~H. Tan, ``Removal of low frequency transient noise from
  old recordings using model-based signal separation techniques,'' in
  \emph{Proceedings of the IEEE Workshop on Applications of Signal Processing
  to Audio and Acoustics (WASPAA 1997)}.\hskip 1em plus 0.5em minus 0.4em\relax
  New Paltz, USA: IEEE, October 1997.

\bibitem{esquef-et-al-long-pulse}
P.~A.~A. Esquef, L.~W.~P. Biscainho, and V.~V{\"a}lim{\"a}ki, ``An efficient
  algorithm for the restoration of audio signals corrupted with low-frequency
  pulses,'' \emph{Journal of the Audio Engineering Society}, vol.~51, no.~6,
  pp. 502--517, 2003.

\bibitem{esquef-welter}
P.~A.~A. Esquef and G.~S. Welter, ``Audio de-thumping using {H}uang's empirical
  mode decomposition,'' in \emph{Proceedings of the 14th. International
  Conference on Digital Audio Effects (DAFx-11)}, Paris, France, September
  2011.

\bibitem{carvalho-avila-biscainho-aesbrasil-2014}
H.~T. Carvalho, F.~R. \'Avila, and L.~W.~P. Biscainho, ``A {B}ayesian procedure
  for restoration of audio signals degraded by low-frequency pulses,'' in
  \emph{Proceedings of the 12th. Brazilian Conference on Audio
  Engineering}.\hskip 1em plus 0.5em minus 0.4em\relax S{\~a}o Paulo, Brazil:
  AES-Brazil, May 2014, pp. 47--54.

\bibitem{esquef-biscainho-diniz}
P.~A.~A. Esquef, L.~W.~P. Biscainho, and P.~S.~R. Diniz, ``Detec\c{c}\~{a}o de
  pulsos longos em sinais de \'{a}udio,'' in \emph{Proceedings of the XVII
  Brazilian Symposium on Telecommunications (SBrT)}.\hskip 1em plus 0.5em minus
  0.4em\relax Vila Velha, Brazil: SBrT, September 1999, pp. 191--196, in
  Portuguese.

\bibitem{bishop}
C.~M. Bishop, \emph{Pattern Recognition and Machine Learning}.\hskip 1em plus
  0.5em minus 0.4em\relax New York, USA: Springer, 2007.

\bibitem{rasmussen-williams}
C.~E. Rasmussen and C.~K.~I. Williams, \emph{Gaussian Processes for Machine
  Learning}, 1st~ed.\hskip 1em plus 0.5em minus 0.4em\relax Cambridge, USA: MIT
  Press, 2006.

\bibitem{mackay}
D.~J.~C. MacKay, \emph{Information Theory, Inference and Learning Algorithms},
  1st~ed.\hskip 1em plus 0.5em minus 0.4em\relax Cambridge, UK: Cambridge
  University Press, 2003.

\bibitem{wilson-phd}
A.~G. Wilson, ``Covariance kernels for fast automatic pattern discovery and
  extrapolation with gaussian processes,'' \uppercase{D}.\uppercase{S}c.
  Thesis, University of Cambridge, Cambridge, UK, 2014.

\bibitem{oppenheim-schafer-sig-prog}
A.~V. Oppenheim and R.~W. Schafer, \emph{Discrete-Time Signal Processing},
  3rd~ed.\hskip 1em plus 0.5em minus 0.4em\relax Upper Saddle River, USA:
  Prentice Hall, 2009.

\bibitem{jain}
A.~K. Jain, \emph{Fundamentals of Digital Image Processing}, 1st~ed.\hskip 1em
  plus 0.5em minus 0.4em\relax Upper Saddle River, USA: Pearson, 1988.

\bibitem{cryer-chan}
J.~D. Cryer and K.-S. Chan, \emph{Time Series Analysis With Applications in R},
  2nd~ed.\hskip 1em plus 0.5em minus 0.4em\relax New York, USA: Springer, 2008.

\bibitem{geman-geman}
S.~Geman and D.~Geman, ``Stochastic relaxation, {G}ibbs distributions, and the
  {B}ayesian restoration of images,'' \emph{IEEE Transactions on Pattern
  Analysis and Machine Intelligence}, vol.~6, no.~6, pp. 721--741, 1984.

\bibitem{robert-casella-small}
C.~Robert and G.~Casella, \emph{Introducing Monte Carlo Methods with R},
  1st~ed.\hskip 1em plus 0.5em minus 0.4em\relax New York, USA: Springer, 2009.

\bibitem{robert-casella-big}
------, \emph{Monte Carlo Statistical Methods}, 2nd~ed.\hskip 1em plus 0.5em
  minus 0.4em\relax New York, USA: Springer, 2005.

\bibitem{givens-hoeting-computational-statistics}
G.~H. Givens and J.~A. Hoeting, \emph{Computational Statistics}, 2nd~ed.\hskip
  1em plus 0.5em minus 0.4em\relax Hoboken, USA: Wiley, 2012.

\bibitem{carvalho-dsc}
H.~T. Carvalho, ``Bayes meets bach: Applications of bayesian statistics to
  audio restoration,'' \uppercase{D}.\uppercase{S}c. Thesis, Universidade
  Federal do Rio de Janeiro, Rio de Janeiro, Brazil, 2017.

\bibitem{itu}
ITU-R, \emph{Rec. BS.1387-1: Method for Objective Measurements of Perceived
  Audio Quality}.\hskip 1em plus 0.5em minus 0.4em\relax Geneva, Switzerland:
  International Telecommunication Union, 2001.

\bibitem{kabal-peaq}
P.~Kabal, ``An examination and interpretation of {ITU-R} {BS}.1387: Perceptual
  evaluation of audio quality,'' McGill University, Montreal, Canada, In:
  Report 217, 2002.

\bibitem{esquef-web}
P.~A.~A. Esquef, ``Paulo esquef - research on digital audio restoration,''
  https://www.lncc.br/\~{}pesquef/AR\_pages.html.

\bibitem{catalan-cylinders}
B.~de~Catalunya, ``Uns incunables del sonor - la col·lecci{\'o}
  regordosa-turull de cilindres de cera,''
  http://www.bnc.cat/Exposicions/Uns-incunables-del-sonor/Continguts-de-l-exposicio.

\end{thebibliography}
\end{document}